\documentclass[12pt]{article}

\usepackage{epsfig}
\usepackage{cite}
\usepackage{amsmath, amssymb, amsfonts}
\usepackage{color}
\usepackage{latexsym}
\usepackage{graphicx}
\usepackage{cancel}
\usepackage[colorlinks,bookmarks]{hyperref}
\hypersetup{pdfpagemode=UseNone, pdfstartview=FitH, linkcolor=blue, citecolor=red, urlcolor=blue}

\def\be{\begin{equation}}
\def\ee{\end{equation}}
\def\ba{\begin{eqnarray}}
\def\ea{\end{eqnarray}}

\def\bdm{\begin{displaymath}}
\def\edm{\end{displaymath}}
\def\la{~\mbox{\raisebox{-.6ex}{$\stackrel{<}{\sim}$}}~}

\def\bq{\begin{quote}}
\def\eq{\end{quote}}

 at 10truept

\newcommand{\de}{\partial}

\renewcommand{\(}{\left(}
\renewcommand{\)}{\right)}

\newcommand{\eps}{\epsilon}

\newcommand{\om}{\omega}
\newcommand{\Om}{\Omega}

\newcommand{\Mpl}{M_{\mathrm{Pl}}}

\newcommand{\bea}{\begin{eqnarray}}
\newcommand{\eea}{\end{eqnarray}}

\newcommand{\bi}{\begin{itemize}}
\newcommand{\ei}{\end{itemize}}

\newcommand{\beq}{\begin{equation}}
\newcommand{\eeq}{\end{equation}}
\newcommand{\beqa}{\begin{eqnarray}}
\newcommand{\eeqa}{\end{eqnarray}}
\newcommand{\mpl}{\Mpl}

\def\la{~\mbox{\raisebox{-.6ex}{$\stackrel{<}{\sim}$}}~}

\def\12{{1 \over 2}}

\def\ltap{\ \raise.3ex\hbox{$<$\kern-.75em\lower1ex\hbox{$\sim$}}\ }
\def\gtap{\ \raise.3ex\hbox{$>$\kern-.75em\lower1ex\hbox{$\sim$}}\ }
\def\gl{\ \raise.5ex\hbox{$>$}\kern-.8em\lower.5ex\hbox{$<$}\ }
\def\roughly#1{\raise.3ex\hbox{$#1$\kern-.75em\lower1ex\hbox{$\sim$}}}

\begin{document}

\thispagestyle{empty}
\begin{flushright}
July 2024 
\end{flushright}
\vspace*{1.1cm}
\begin{center}

{\Large \bf Reviving QFT in $2+1$ de Sitter Spacetime}

\vspace*{1cm} {\large Guido D'Amico$^{a, }$\footnote{\tt damico.guido@gmail.com} and 
Nemanja Kaloper$^{b, }$\footnote{\tt
kaloper@physics.ucdavis.edu}
}\\
\vspace{.5cm} 
{\em $^a$Dipartimento di SMFI dell' Università di Parma and INFN}\\
\vspace{.05cm}{\em Gruppo Collegato di Parma, Italy}\\
\vspace{.3cm}
{\em $^b$QMAP, Department of Physics and Astronomy, University of
California}\\
\vspace{.05cm}
{\em Davis, CA 95616, USA}\\

\vspace{1.2cm} ABSTRACT
\end{center}
We consider a conformally coupled scalar QFT on $2+1$ dimensional static Einstein 
universe $R \times S^2$, and write down the free theory Hilbert space. We explain that this theory is 
secretly a QFT in $2+1$ de Sitter space because all the quantum observables experience {\it quantum revivals}, 
which naturally restricts the timelike $R$ to the appropriate de Sitter time range. 
Our construction circumvents the 
causal obstruction to formulating QFT in de Sitter due to event horizons. There aren't any in static Einstein. 
The `unitary gauge' description of the theory is realized by the zonal harmonics $P_\ell(\hat n \cdot \hat n')$.
We verify that interactions with conformally invariant 
external sources are mediated only by these modes. 
Hence these modes comprise the complete basis of the ``bulk" theory. 
When the theory is cut off in the UV the basis dimension scales as the Bekenstein-Hawking formula. 

\vfill \setcounter{page}{0} \setcounter{footnote}{0}

\vspace{1cm}
\newpage

\section{Prologue}

A first-principles formulation of quantum field theory (QFT) in de Sitter space remains an open and 
very challenging problem 
\cite{Banks:2000fe,Banks:2001yp,Hellerman:2001yi,Fischler:2001yj,Witten:2001kn,Strominger:2001pn,Spradlin:2001pw,Dyson:2002pf}. 
Even if we ignore gravity, for example by decoupling it in the limit $\mpl \rightarrow \infty$, de Sitter space still features horizons
which obstruct defining ``in" and ``out" QFT states and 
the conventional Minkowski space approach to defining the $S$-matrix. 
If one attempts to ignore the horizon, by formulating a QFT on only one side of it, one 
encounters a plethora of confusing puzzles, such as 
vacuum ambiguities \cite{Mottola:1984ar,Allen:1985ux}, imaginary contributions
to mode phases which lead to a mode ``freezeout" at large wavelengths and suggest a failure of unitarity \cite{Strominger:2001pn}, 
negative energies from superhorizon modes \cite{Abbott:1981ff}, 
and so on. 

It is very puzzling that some of these features appear to have significant applications to the real 
world physics, such as the genesis of structures in cosmology \cite{Mukhanov:1981xt}, despite the lack of, as yet, a clear complete 
definition of the underlying principles (for some insights, see \cite{Witten:2021jzq}). Thus it is clearly
important to better understand QFT in de Sitter. 

Here we endeavor to shed some more light on the issue of QFT in de Sitter, by pursuing a curious clue which comes from the behavior 
of worldlines of null observers. de Sitter space appears infinite to any timelike observer  in any of its complete coordinate atlases. However
null observers which semi-circumnavigate de Sitter do so in finite affine time. 
An example is the light-rays traveling from the North Pole to the South Pole. A timelike observer takes an infinite proper time 
to see this, but this divergence is irrelevant for null probes.  

This highlights a peculiar feature of conformal observers, who traverse de Sitter in finite range of affine parameter, that measures 
the ``causal size" of spacetime. Such observers could be extended past the spacelike boundaries. 
Indeed, a conformal field theory on de Sitter actually dwells on a family of spacetimes which are conformally related to de Sitter, and 
which include the static Einstein universe $R \times S^2$. This space is not bounded in time by spacelike boundaries at finite range of 
affine parameter. As a result it does not have horizons. 

Therefore a quadratic (i.e., ``free") QFT can be straightforwardly formulated on the static Einstein 
space $R \times S^2$, only accounting for subtleties with compact base manifold. 
Despite the apparent simplicity of the framework, there are interesting and nontrivial features which are revealed. Perhaps most notably, 
since the spectrum of the theory is discrete, and the energy levels are odd integer multiples of the lowest energy level, the 
quantum evolution undergoes quantum revivals \cite{Jaynes:1963zz,Eberly:1980zz}\footnote{For examples of revivals in QFT see \cite{Cardy:2016lei,Dowker:2016twk}.}. This implies the theory is dynamically restricted in \underbar{\it time} back to de Sitter space. 
The evolution beyond this time interval, in its past or its future, is a trivial repetition of ``already seen". 
In other words, a conformally coupled scalar in static Einstein $R \times S^2$ is secretly a QFT in de Sitter. 

Thanks to this feature, static Einstein space, in a role of a covering space of de Sitter, becomes a useful proxy 
to define a QFT on de Sitter. Once the horizons are removed, the procedure for constructing the Hilbert space
is straightforward. The theory lives on a static sphere, and we can quantize it. 
The spectrum is discrete, gapped by the radius of the sphere $\rho$. 
We also outline the flat space limit of the quadratic theory on $R \times S^2$ 
using In\"on\"u-Wigner contraction \cite{Inonu:1953sp}. When the $S^2$ flattens out, we get two flat space scale-invariant 
continuum QFTs, as images of the hemispherical restrictions of the initial QFT on static Einstein. They also correspond to the limit 
of two antipodal static patches of de Sitter in the flat space limit $H_0 \rightarrow 0$. In this case, the spectra become continuous,
since the gap vanishes, and the spatial sections decompactify. 
In this sense, the limit is singular\footnote{If this limit were 
a different conformal gauge of the theory, then many of the continuum of states should be pure gauge. 
A construction of the correspondence between observables on finite $S^2$ and in infinite radius limit 
might be able to test this. However will not delve into this interesting question here.}.

The standard Minkowski space formulation of interactions in quantum mechanics and in QFT, 
where one introduces couplings  using the adiabatic switching on and off procedure (e.g. see \cite{messiah}) 
to connect the theory to the free quadratic limit in infinite past and future does not seem to directly apply here. If we 
restrict the evolution to $I \times S^2$, we do not seem to be able to adiabatically switch on and off the interactions, since the conformal 
time span is finite. 
If on the other hand we try to formulate interactions on the full static Einstein $R \times S^2$, switching the interactions 
on and off as $\eta \rightarrow \pm \infty$ violates the periodicity of the revivals, which raises questions about interpreting the theory as
secretly living on de Sitter, the order of limits, and even the applicability of the free theory as the starting point. 
Due to these challenges, we set aside the problem of formulating real self-interactions for now. 

Instead, as a proxy for interactions
we consider here external sources, and how they interact via the scalar field exchange. 
The picture which emerges is that sources interact with each other
pairwise, as is common, with the interaction mediated by the conformally coupled scalar.
The two-point scalar function\footnote{The subscript denotes that the theory is free.} $\langle  \phi(x) \phi(x') \rangle_0 = ~_0\langle 0 | \phi(x) \phi(x') | 0 \rangle_0$ which mediates the momentum transfer between the sources can be resummed, and the result is a series in precisely the nonlocal modes $P_\ell(\hat n \cdot \hat n')$.
In other words, the virtual quanta which yield momentum transfer between sources are just the bulk terms that enshrine the 
minimal set of degrees of freedom and no more. The local bulk modes $Y_{\ell m}$ merely encode various externally selected 
reference frames introduced by an observer, but how they propagate is restricted by symmetry to only involve
$P_\ell$'s. 

Without the cutoff, the Hilbert space of the theory is countably infinite. However if we introduce a covariant 
UV-cutoff, by e.g. adding to the theory the minimal gravity in $2+1$, which has no local gravitons 
\cite{Deser:1983tn,Deser:1983nh,Achucarro:1986uwr,Witten:1988hc}, but introduces the dimensional Newton's constant,
and so a scale $M_3=1/G_3$, the dimension of the Hilbert space becomes finite. The reason is that when $2+1$ gravity is 
present, the dimensional sources influence geometry by sourcing a deficit angle, which locally requires that
no masses can exceed the critical value $M_{crit} = 1/(4 G_3)$ \cite{Deser:1983tn,Deser:1983nh}. 
Introducing the UV cutoff may seem concerning 
because it may disturb the symmetries of the scalar sector. Yet, a conceptually similar UV regularization via gravity in $3+1$ dimensions, 
which was treated as a spectator, was used by 't Hooft in the brick wall formulation of black hole entropy \cite{tHooft:1984kcu}, 
ushering the holographic approach to formulating quantum gravity. 

As we noted, on a finite $S^2$, the local eigenmodes $Y_{\ell m}(\hat n)$ of the QFT are restricted by symmetry to always 
add up to $P_\ell$'s by spherical harmonics addition theorem. Hence $2\ell+1$ local 
modes at each frequency level only propagate as a single nonlocal degree of freedom $P_\ell$. The number of 
these nonlocal modes, after imposing the UV bound as outlined above, is ${\#_{\tt Hilbert}} \sim \sum^{\ell_{crit}}_{\ell=0} 1 \sim \ell_{crit} \la  \frac{1}{G_3H_0}$. 
It scales precisely as the proper perimeter of the largest circle on $S^2$ -- the equator -- which replaces the 
horizon area in $2+1$ de Sitter. The total number of $P_\ell$ modes saturates the Bekenstein-Hawking bound. 
This agrees well with the ideas about the dimension of the Hilbert space of a QFT in
de Sitter being bounded by the Bekenstein-Hawking formula \cite{Banks:2000fe,Banks:2001yp}. 

In the next section, we give a concise summary of the conceptual points of this paper. 
Following it, we provide the details, and in the Summary we 
highlight our results and their physical implications. 

\section{A Prolegomenon to the Paper}

In this work, we will explore QFT in de Sitter in $2+1$ dimensions since it provides mathematical simplicity. 
Similar explorations could be done in de Sitter in more dimensions. We will also mostly 
ignore gravity by dynamically decoupling it, taking $2+1$ Planck scale $M_3 \rightarrow \infty$ when required; in the end, we will restore it 
as a spectator, in order to use $M_3$ as a UV regulator. While this breaks conformal symmetry explicitly, the absence of perturbative 
gravitons in $2+1$ makes it plausible that the treatment of the QFT as a perturbative conformal field theory is not a terrible 
approximation. Bearing this subtlety in mind, this means we will work with the fixed $dS_{2+1}$ metric, which 
in global coordinates is
\be
d{\cal S}^2 = - dt^2 + \frac{\cosh^2(H_0 t)}{H_0^2} \Bigl(d\theta^2 + \sin^2(\theta) \, d\chi^2 \Bigr) \, .
\label{dsmet}
\ee
The angular part $d\Omega_2  = d\theta^2 + \sin^2(\theta) \, d\chi^2$ 
is the metric on the unit sphere $S^2$. The scale of the physical radius is set by the de Sitter Hubble parameter, $\rho = 1/H_0$. 
This metric is invariant under the group of de Sitter isometries $SO(3,1)$, 
which has six generators in $2+1$. 

\begin{figure}[thb]
    \centering
    \includegraphics[width=8cm]{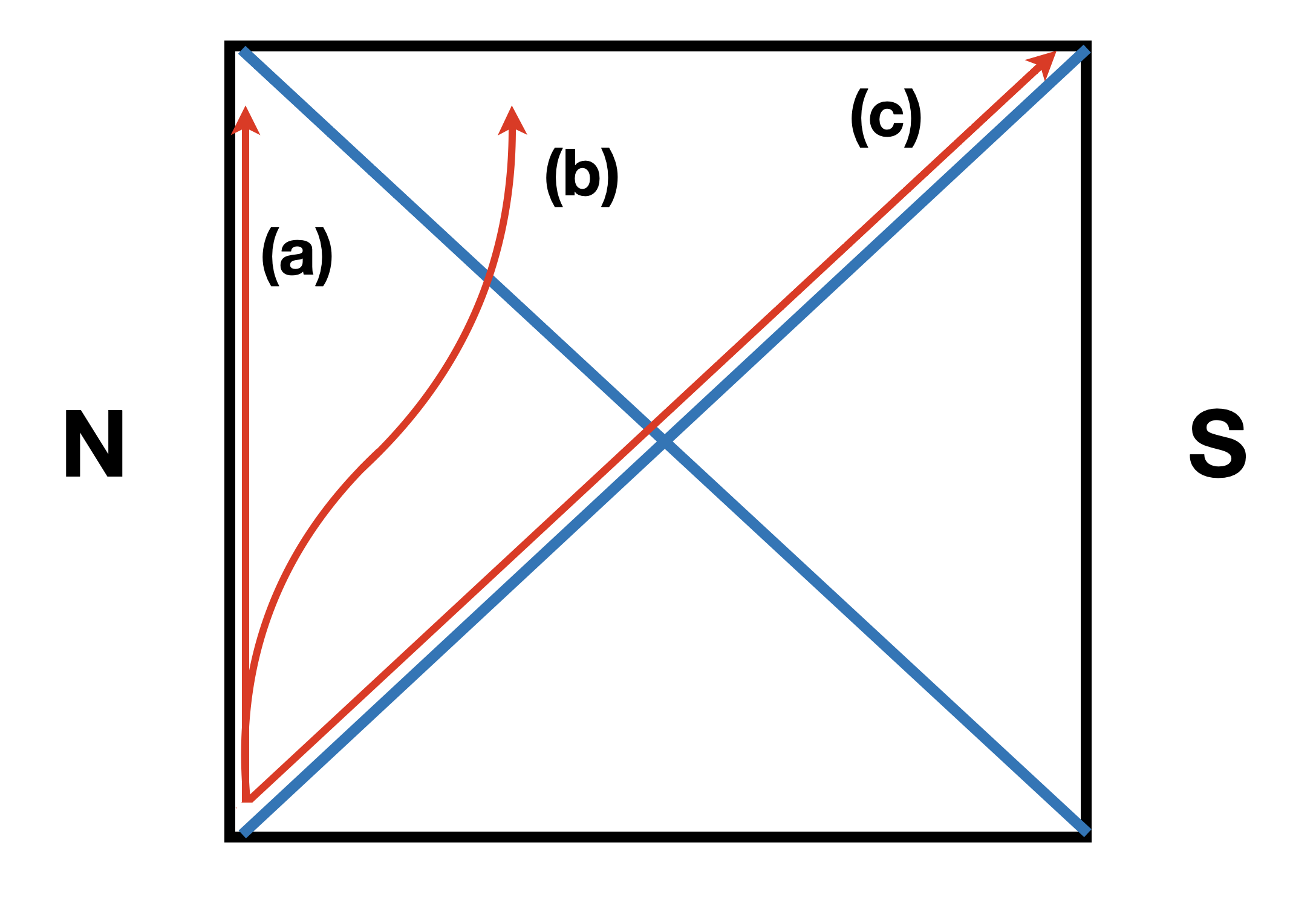} 
     \caption{Causal geodesics in de Sitter: (a) a rest frame timelike geodesic; (b) a radial timelike geodesic; (c) a null geodesic. }
    \label{fig1}
\end{figure}

At face value, this line element metricizes a spacetime with a geometry of a spatial sphere $S^2$, with an evolving radius,
which starts out at infinity, shrinks for an infinite range of de Sitter global time $t \in (-\infty,0)$, and then ``bounces" from the minimal
radius $\rho = 1/H_0$ at $t=0$ and re-expands back to infinite size at $t \rightarrow \infty$. This suggests that the spacetime
``starts" and ``ends" at $t=\mp \infty$, respectively. However, there is a subtlety; to see it, 
let's consider the worldlines of causal observers; i.e., timelike and null geodesics in de Sitter. For simplicity,
we ignore the geodesics along which $\chi$ varies, and only consider smooth great 
circles on $S^2$. For timelike geodesics, in principle we can always 
go to the observer's rest frame, where $\theta = {\rm const}$, and so the proper time 
along the geodesic is $\Delta \tau = \int \sqrt{-d{\cal S}^2} = \Delta t$. This means that a timelike geodesic which initiates 
from the past spacelike boundary of de Sitter and extends to the future spacelike boundary takes up an infinite proper time
to ``arrive". The same remains true from the vantage point of other observers, in their own rest frames. If we consider 
a timelike geodesic along which $\theta$ changes, such that 
\be
\frac{d\theta}{dt} = \frac{c H_0}{\cosh(H_0 t) \sqrt{\cosh^2(H_0 t) + c^2}} \, ,
\ee
where $c \ne 0$ is a dimensionless integration constant, we find that for geodesics that originate and terminate close to the
spacelike boundaries, as $t \rightarrow \pm \infty$, 
\be
\tau = \int dt \frac{\cosh(H_0 t)}{\sqrt{\cosh^2(H_0 t) + c^2}} \sim t \rightarrow \pm \infty \, .
\ee
Therefore, regardless of who is clocking the time, the spacelike boundaries of de Sitter are separated by a divergent 
proper time interval, depicted by cases (a) and (b) in Fig. (\ref{fig1}). 

\begin{figure}[thb]
    \centering
    \includegraphics[width=11cm]{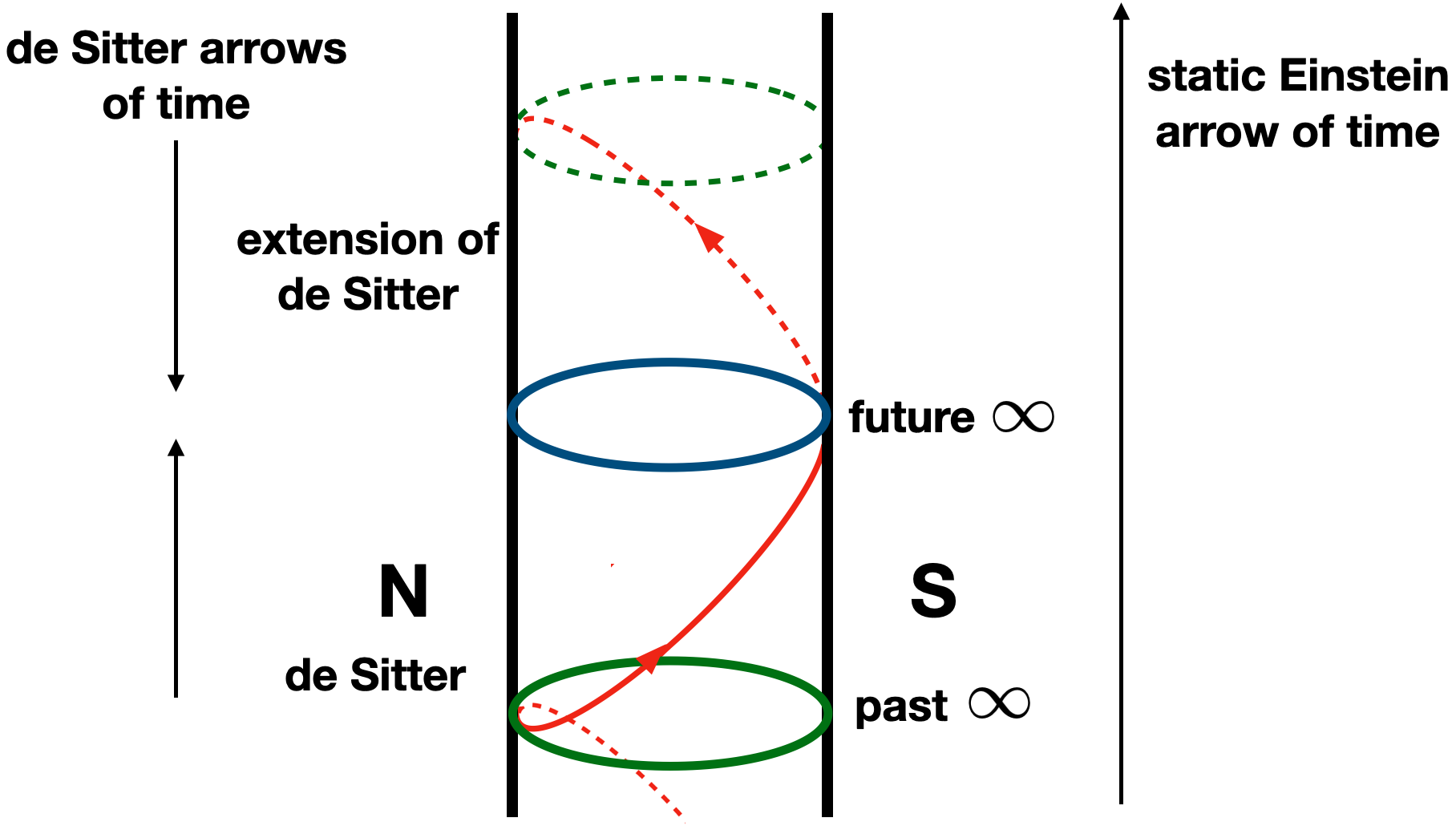} 
     \caption{A null geodesic extending beyond spacelike future and past infinities.}
    \label{fig2}
\end{figure}

Curiously (and amusingly) this completely changes when we turn to null geodesics. 
For null geodesics, $dt = \pm d\theta \cosh(H_0 t)/H_0$. If we pick 
a future oriented null geodesic going outward from the North Pole of $S^2$ (depicted by (c) in Fig. (\ref{fig1})), and use the dimensionless conformal
time $d \hat \eta = H_0 dt/\cosh(H_0 t)$, this equation becomes $d\hat \eta = d\theta$. Thus this null geodesic reaches the South Pole of $S^2$
in a finite interval of the dimensionless conformal time $\Delta \hat \eta = \pi$. Since we can pick the affine parameter along this geodesic to be
precisely the conformal time, as the geodesic equations retain 
their minimal form due to $g_{\mu\nu} \frac{d x^\mu}{d\hat \eta} \frac{d x^\nu}{d\hat \eta} = 0$
(see Appendix D of \cite{Wald:1984rg}),
\be
\frac{d^2 x^\mu}{d\hat \eta^2} + \hat \Gamma^\mu_{\nu \lambda} 
\frac{d x^\nu}{d\hat \eta} \frac{d x^\lambda}{d\hat \eta} = 0 \, , 
\label{geodeq}
\ee
this means that the null geodesics
reach their antipodal point on the sphere after a finite affine parameter interval 
$\Delta \hat \eta = \pi$, having started at exactly the past spacelike boundary 
of de Sitter and ending at exactly the future spacelike boundary. Hence ``photons" 
which follow these geodesics take only a finite ``time" to travel from de Sitter past to de Sitter future 
boundary, semi-circumnavigating the sphere. Since the geodesics are null, their tangent vectors at the 
boundaries do not vanish, but are indeterminate. Therefore we can in principle extend these geodesics beyond the future 
boundary, as in Fig. (\ref{fig2}) - or not, in which case we'd have to declare what happens to such a geodesic when it 
reaches the relevant ``infinity". As explained by Wald \cite{Wald:1984rg} (Sec. 9.1, p. 216) such situations are 
pathological, which prompts providing additional information about what might happen there.

In this work we will address this ambiguity by extending null geodesics beyond spacelike de Sitter boundaries in a 
conformally related geometric frame effectively defining ``covering space" of de Sitter. We will work with conformally 
coupled scalars whose wave packets follow the null geodesics. Such probes reach the standard de Sitter
spacelike boundaries in a finite affine parameter range. For them, the ``end of time" boundaries are absent, and we will 
ignore them because there are no intrinsic non-null probes of the spacetime. The emergent covering space, depicted in 
Fig. (\ref{fig2}), is the family of static Einstein universes in $2+1$, with topology $R \times S^2$.  As noted, the radius of 
$S^2$ is a pure gauge parameter, which can be changed at will by rigid conformal transformations. A specific choice
for $R \times S^2$ corresponds to fixing the theory to a specific element of the scaling symmetry orbit. 

Using this picture as a motivation, we can formulate in detail a conformally coupled scalar field theory $\phi$ on static Einstein 
universe $R \times S^2$. 
We will write down the complete quadratic theory Hilbert space, quantizing the theory using the Hamiltonian eigenmodes, which 
comprise a discrete system of harmonic oscillators, with positive and negative frequencies. The spectrum is discrete, 
because the QFT lives on a spatially compact space $S^2$ with a conformal gauge-fixed sphere radius $\rho = 1/H_0$. 
The positive and negative frequencies are separated from each other by a gap, 
and the time evolution of all the states, clocked by the static Einstein time, 
is realized by phase shifts with constant frequencies 
just like in flat space. Thus the vacuum definition is globally valid: 
there is no Bogoliubov particle production\footnote{This was also noted in \cite{Lagogiannis:2011st,Bousso:2001mw} for the vacuum. In 
fact in our case this is true for all the states we consider, regardless of whether we use the de Sitter conformal gauge or
the static Einstein conformal gauge.}. The same extends to all the Hamiltonian eigenstates with a fixed
excitation occupation number, however. 

The mode frequencies are odd integers, separated by $\Delta E = 1/\rho = H_0$ on any $R \times S^2$ of 
a fixed sphere radius $\rho = 1/H_0$. Hence time evolution of an arbitrary state is given by the linear superposition 
of time-translated Hamiltonian eigenstates, whose time translation is merely a phase shift. Therefore
it turns out that time evolution is {\it exactly periodic}: any wavefunction and any observable 
defined by an expectation value of an operator in a given state experiences a quantum revival
\cite{Jaynes:1963zz,Eberly:1980zz} after a finite time interval. 
Because the frequency spectrum is a set of odd integers, 
the revival time is completely independent of the wavefunction. Hence the time axis $R$ is dynamically 
restricted to an interval\footnote{The interval, rather than a circle,
arises because the periodicity of the observables, which arises from combining time evolution and parity, 
on the interval does not require periodicity of the wavefunction with the {\it same} period. The observables, defined as operator matrix elements, are periodic with a shorter period than wavefunctions.}. 
Beyond this fundamental interval, all observables repeat regularly for any choice of
the initial state. 

Therefore the total evolution of all observables is completely described 
by evolution within a fixed compact time interval, and so we can formally restrict $R \times S^2$ to $I \times S^2$, reintroducing
the past and future time boundaries. Although the spacetime $I \times S^2$ has fewer isometries (it has 4 Killing vectors) than de Sitter in
three dimensions (which, as noted, 
has 6 Killing vectors), two extra dynamical isometries arise due to the conformal
symmetry of the scalar on $I \times S^2$. This restores the isomorphism between the dynamics on these two spacetimes. 

Curiously, using conformal transformations and In\"on\"u-Wigner contraction \cite{Inonu:1953sp}, 
any such discrete conformal gauge-fixed theory can be mapped onto
the continuum limit, as long as conformal symmetry remains unbroken\footnote{Even quantum-mechanically, as there do not appear to be conformal anomalies in $2+1$ \cite{Dowker:1978md,Christensen:1978yd,Birrell:1982ix,Cappelli:1988vw}.}. The spectra become continuous
because the spatial sections decompactify. This maps de Sitter horizons to 
Minkowski null infinites, and decouples the theory into two separate copies of the scale-invariant 
scalar field theory in $2+1$-dimensional Minkowski. Comparing the flat space limit to the theory on $R \times S^2$ 
with a finite radius of $S^2$ is tricky due to the spectrum becoming continuous. The continuum limit
effectively changes the boundary conditions of modes on $S^2$ to those on $R^2$, introducing the
``boundary" at infinity. We will set aside this interesting question here. 

Still, even the usual counting of modes on discretely infinite Hilbert space on a $R \times S^2$ involves redundant degrees 
of freedom, since there are
additional symmetries which relate them. Those are precisely the residual $S^2$ rotations, which 
transform any two antipodal points on $S^2$ into any other two, without 
changing the base space. Using the spherical harmonics {\it addition theorem} (see, e.g. \cite{Bander:1965rz,Bander:1965im})
\be
P_\ell(\hat n \cdot \hat n') = \frac{4\pi}{2\ell+1} \sum_{m=-\ell}^\ell Y_{\ell m}(\hat n) Y^*_{\ell m}(\hat n') \, , 
\label{sphaddth}
\ee
the Fourier series for functions on $S^2$ is 
\be
\Psi(\hat n) = \sum_{\ell,m} Y_{\ell m}(\hat n) \int_{S^2} d \hat n' \, Y^*_{\ell m}(\hat n') \Psi(\hat n') = 
\frac{2\ell+1}{4\pi} \sum_\ell \int_{S^2} d \hat n' \, P_\ell(\hat n \cdot \hat n') \Psi(\hat n') \, ,
\label{fourier}
\ee
where $d\hat n$ is the invariant measure on $S^2$. From this equation, we see that the local basis of 
$2\ell+1$ functions for each $\ell$ can be faithfully traded for a single nonlocal function 
$P_\ell(\hat n \cdot \hat n')$ for each $\ell$. The 
nonlocality of the argument allows us to reduce the physical Hilbert space at any specific point 
to the span of only zonal spherical harmonics $P_\ell$. The point here is that the additional
$2\ell$ modes $Y_{\ell m}$ are needed when we fix one observer as an `accountant' who records
the observations of all other observers on $S^2$. When we restore the equality of all observers 
$P_\ell$'s suffice.

Although we cannot confirm that this persists in a fully nonlinear interacting theory, we can at least
check that the interactions of external sources mediated by the conformally coupled scalar exchange
maintain this feature. Since the theory is bilinear, we 
can solve for their interactions exactly. 
The source-source amplitude, which is equal to the scalar two-point 
scalar function $\langle  \phi(x) \phi(x') \rangle_0 = ~_0\langle 0 | \phi(x) \phi(x') | 0 \rangle_0$, is a series in the nonlocal 
modes $P_\ell$, after partial resummation over the polar quantum number $m$. This means that only those modes which 
can be associated with the nonlocal modes basis describe  
the virtual quanta which control momentum transfer between the sources. In this sense, the conformally gauge-fixed 
theory on $I \times S^2$ is analogous to a unitary gauge.

After transitioning to the nonlocal basis, the resulting Hilbert space is still 
discretely infinite in the conformal limit. However if a UV regulator is introduced, for example by adding the (topological) $2+1$ gravity 
\cite{Deser:1983tn,Deser:1983nh,Achucarro:1986uwr,Witten:1988hc} 
which brings in the $2+1$ Planck length, the Hilbert space of the QFT is cut off in the UV. 
When this happens, the Hilbert space 
dimension is finite, and it scales as the proper perimeter of the largest circle on $S^2$ -- the equator -- which is the apparent 
horizon of the static Einstein $R \times S^2$. It's perimeter is the same as the 
horizon perimeter in $2+1$ de Sitter: the stationary apparent horizon of static Einstein coincides with the de Sitter
horizon at the horizon bifurcation point, representing de Sitter bounce.  This implies that the 
Hilbert space dimension is bounded by the Bekenstein-Hawking formula, from either the
static Einstein or the de Sitter vantage point.  

We stress again that a UV regularization of a conformal QFT may raise concerns about the role of conformal symmetry, 
and we do not have a complete picture of it here. However, we take a reassurance 
from the quantitative success of a similar application 
of gravity as a UV regulator in the case of brick wall formulation of black hole entropy \cite{tHooft:1984kcu}, 
which may coexist with emergent conformality in near-horizon limits. Thus we temporarily set these concerns aside 
and consider the UV regulator only in the leading order. It would be interesting to understand this issue better.

In the remainder of the paper we turn to putting some flesh on these bones. 

\section{Conformally Coupled Scalar on 2+1 Static Einstein}

We consider a conformally coupled scalar field on a fixed $2+1$-dimensional background metric, so that
the field theory is governed by the action 
\be 
S = - \int d^3x \sqrt{g}  \Bigl( \frac12 \de_\mu \phi \de^\mu \phi + \frac{1}{16} R \phi^2 
+ \frac{1}{6!} g_6 \phi^6 - \phi J \Bigr) \, .
\label{action}
\ee
The parameter $g_6$ is the dimensionless scalar coupling constant. Here we have included external sources $J$. 
This theory is clearly scale invariant. However, 
the conformal coupling $\propto R \phi^2$ ensures that the theory is invariant under general local 
conformal transformations \cite{Wald:1984rg},
\ba
&&\phi(x) \rightarrow \bar \phi(x)= \frac{1}{\sqrt{\Omega(x)}} \, \phi(x) \, , ~~~~~~~~~~ 
g_{\mu\nu}(x) \rightarrow \bar g_{\mu\nu}(x) = \Omega^2(x) g_{\mu\nu}(x) \, , \nonumber \\
&& ~~~~~~~~~~~~~~~~~~~~~~ 
J(x) \rightarrow \bar J(x)= \frac{1}{{\Omega^{5/2}(x)}} \, J(x)\, .
\label{weyl}
\ea
Here, $\Omega$  is a completely arbitrary scalar function. 

Our ultimate 
purpose here is to explore QFT in de Sitter. We study it in the single coordinate chart  of global coordinates, 
which cover de Sitter faithfully. In these coordinates, the 
de Sitter metric is given by Eq. (\ref{dsmet}). Since the theory is  conformally invariant, we can 
in fact consider the theory on any metric conformal to  (\ref{dsmet}), and as long as the conformal transformation
is smooth and non-singular the information obtained must be the same. 

A very convenient conformal image of de Sitter is obtained by embedding it into static Einstein $R \times S^2$. 
This is realized by straightforwardly replacing the global time $t$  by the conformal time 
\be
d\eta = \frac{dt}{\cosh(H_0 t)} \, ,
\label{conftime}
\ee
which after integration yields
\be
\cosh(H_0 t) = - \frac{1}{\sin(H_0 \eta)} \, .
\label{comconftime}
\ee
The minus sign ensures that both the global de Sitter time $t$ 
and the globally defined conformal time $\eta$ of $R \times S^2$
share the same arrow of time in the region $\eta \in (-\pi/H_0,0^-)$, reflecting the standard convention for parameterizing
conformal time in approaches to inflation which use flat slicing charts of de Sitter, valid near its future spacelike boundary.
In terms of the time $\eta$, the metric (\ref{dsmet}) becomes, using $\rho = 1/H_0$,
\be
 d {\cal S}^2 = \frac{1}{\sin^2(\frac{\eta}{\rho})} \Bigl(-d\eta^2 + \rho^2 d\Omega_2 \Bigr) \, .
\label{confds}
\ee
Note that $d \Om_2 = d\theta^2  + \sin^2(\theta) d\chi^2$ is the metric on the unit $S^2$.

Since we are pursuing a conformally invariant scalar field theory (\ref{action}) on this background, we can 
absorb the conformal warp factor in (\ref{confds}) by the Weyl transformation (\ref{weyl}) using 
\be
\Omega = -\sin\(\frac{\eta}{\rho}\) \, ,
\label{weyltrafo}
\ee
which is positive definite in the interval $\eta \in (-\pi \rho,0^-)$, and map the theory (\ref{action})
onto a section of the static Einstein $R \times S^2$, with the metric $ds^2 = \Omega^2 d{\cal S}^2$, 
\be 
d s^2 = -  d \eta^2 + \rho^2 d \Om_2 \, .
\label{metricsE}
\ee

On this background, the general field equation which follows from (\ref{action}), 
restricting to $g_6 = 0$ and sources temporarily switched off, $J=0$, and using 
$\square_3$ to denote the d'Alembertian on the static Einstein metric (\ref{metricsE}), is
\be
\square_3 \phi - \frac{R}{8} \phi  = 0 \, ,
\label{fullfeq}
\ee
It can be easily solved. In particular, on the background  (\ref{metricsE}) it becomes
\be
 - \de_\eta^2 \phi + \frac{1}{\rho^2} \nabla_\Om^2 \phi - \frac{1}{4 \rho^2} \phi = 0 \, ,
 \label{freefeq}
\ee
where $\nabla_\Om^2$ is the Laplacian on $S^2$. Clearly, this equation is exactly solvable using separation
of variables on $R \times S^2$. We will present the details momentarily.

Before we proceed with quantizing the conformal-gauge-fixed theory, it is important to address the question
of what happens when evolution in time extends beyond the ``principal interval" $\eta \in (-\pi \rho,0^-)$. 
The metric
(\ref{metricsE}), and the field equation (\ref{freefeq}) are manifestly smooth
as $\eta$ approaches the end points of the interval $I = (-\pi \rho,0^-)$, and the segment $I \times S^2$ is clearly not geodesically 
complete. In particular, there are null geodesics such as those depicted in Fig. (\ref{fig2}) which come in and out of this
segment. 

On the other hand, the conformal transformation (\ref{weyl}), (\ref{weyltrafo}) is singular at the end points, since the 
conformal factor $\Omega$ vanishes there. As $\eta$ approaches the end points, the physical radius of
the $S^2$ factor in de Sitter metric (\ref{dsmet}) blows up uniformly, while the de Sitter time is infinitely dilated relative to the
conformal time. So the extension beyond the interval $I = (-\pi \rho,0^-)$ is perfectly reasonable on static Einstein $R \times S^2$, but it poses questions for the 
pullback of the theory on de Sitter. 

Let's ignore for the moment the singularity of (\ref{weyl}), (\ref{weyltrafo}) at the interval end points. It then appears 
suggestive to think of static Einstein as a cover of an infinite stack of de Sitter geometries, separated by spacelike 
(non-curvature) singularities separated by $\Delta \eta_* = \pi \rho$, where $\Omega$ of (\ref{weyltrafo}) vanishes. Mathematically, this could be 
accomplished, for example, by extending $\Omega$ of (\ref{weyltrafo}) beyond $\eta \in (-\pi \rho,0^-)$ by taking
$\tilde \Omega = |\Om| = | \sin(\eta/\rho)|$, suggested e.g. by the metric transformation in (\ref{weyl}) which only depends on
$\Omega^2$. 

However since this continuation is along the time direction, affecting the phase of the transformed field 
$\phi$ in (\ref{weyl}), and the time evolution of the system with exact free eigenfunctions of (\ref{freefeq}) is in fact a phase
evolution, we must proceed with care. The point is that replacing $\Omega$ by $|\Omega|$ in (\ref{weyl}) changes the 
analytic properties of the $\phi$ propagation eigenfunctions in (\ref{weyl}) around the singularities of $\Omega$, which would prompt the 
modification of the field equations governing $\phi$. Additional singularities would arise from replacing $\Omega \rightarrow |\Omega|$,  
behaving as sources on the spacelike boundaries. To avoid this, we follow the procedure where we
extend the solutions through the end points of $I$ on the static Einstein side without modifying $\Omega$ at 
the end points of the interval $I$, and define their images on de Sitter side
by the inversion of (\ref{weyl}), using the principal value limits of $\Omega$ as $\eta$ approaches the end points on either side. 
Since $\Omega$ given in (\ref{weyltrafo}) has only simple zeros, and so its inverse has simple poles, at the end points, this 
extension will preserve the field equations. 

Enforcing this extension, we can rewrite Eq. (\ref{conftime}), using (\ref{weyltrafo}). as 
\be
dt=  \cosh\(\frac{t}{\rho}\) \, d\eta = - \frac{d\eta}{\sin(\frac{\eta}{\rho})} \, , 
\label{teta}
\ee
which implies that the arrow of time {\it flips} between individual sections of de Sitter covered by static Einstein each time
$\eta$ passes through an interval end point (i.e., crosses a de Sitter spacelike boundary). E.g. if 
$\eta$ passes through zero, for $\eta \rightarrow 0^-$ $dt$ in (\ref{teta}) is {\it positive}. On the other hand, 
for $\eta \rightarrow 0^+$, $dt$ is {\it negative}. Thus from the point of view 
of a hypothetical conventional de Sitter observer who might dare to cross over an interval end point
the de Sitter global time would start to run in the opposite direction upon each crossing. 

Yet, once theory is gauge-fixed to a specific static Einstein, with a fixed $S^2$ radius $\rho$, the conformal time $\eta$ is 
a good smooth variable. It is legitimate to follow evolution parameterized by it, generated by the Hamiltonian
dual to $\eta$, which we are about to derive below. 

With the above pathology affecting 
de Sitter time evolution in mind, it is interesting to ask if the theory
formulated on static Einstein has anything in its structure which reflects the presence of this pathology that is manifest in
other variables. We will see that the answer to this is affirmative, since the theory features exactly 
periodic evolution in the 
conformal time $\eta$, with \underline{all} observables being quantum-mechanically revived. This effectively restricts the conformal 
time variable from $R$ back to the open interval $I$ which covers de Sitter in conformal time. 
This means, that the conformally coupled scalar field on static Einstein is secretly dwelling on de Sitter
restriction of $R \times S^2$.

\section{Quantizing the Theory}

Now we turn to solving the theory (\ref{action}) on a background metricized by (\ref{metricsE}). Since the spatial sections
are $S^2$, it is clear that we will obtain the complete spatial basis of eigenfunctions by expanding in 
spherical harmonics $Y_{\ell m}$. Substituting 
$\phi(x^\mu) = Y_{\ell m}(\hat n) v_{\ell m}(\eta)$ into (\ref{freefeq}) yields, using $\partial_\eta \phi = \dot \phi$ for short,
\be
\ddot v_{\ell m} = - \frac{\ell (\ell + 1) + \frac{1}{4}}{\rho^2} \, v_{\ell m} = - \frac{(2 \ell+1)^2}{4 \rho^2} \, v_{\ell m} \, .
\label{modeeq}
\ee
Therefore the eigenvalues are 
\be
 \om_{\ell m} = \pm \om_{\ell} \, , ~~~~~~ \om_{\ell} = \frac{2\ell +	1}{2\rho} \, ,
 \label{eigenv}
\ee
each being $2\ell+1$ times degenerate. The standard field operator expansion is
\be
  \phi(x^\mu) = \sum_{\ell m} \Bigl( v_{\ell m}(x^\mu) a_{\ell m} + v_{\ell m}^*(x^\mu) a^\dagger_{\ell m} \Bigr) \, ,
  \label{fop}
\ee
where $a_{\ell m}$, $a_{\ell m}^\dag$ are the standard annihilation and creation operators, and $v_{\ell m}$ are the mode functions
$v_{\ell m}(x^\mu) = {\cal A}_{\ell m} Y_{\ell m}(\hat n) e^{- i \om_{\ell} \eta}$.
Here ${\cal A}_{\ell m}$ is fixed by requiring that we recover the canonical commutation relations, 
and the asterisk denotes complex conjugation.

We use the usual equal-time canonical commutation relations in coordinate domain, which are
\be
[ \phi(\vec x_2, \eta), \pi(\vec y_2, \eta)] = i \delta^{(2)}(\vec x_2 - \vec y_2) \, , ~~~~~~~~ 
[ \phi(\vec x_2, \eta), \phi(\vec y_2, \eta)] = [ \pi(\vec x_2, \eta), \pi(\vec y_2, \eta)] = 0 \, .
\label{eqticommrel}
\ee
From the action (\ref{action}), the canonical momentum is 
$\pi(\vec x, \eta) = \frac{\delta S}{\delta \dot \phi} = \rho^2 \sqrt{\gamma_2} \dot \phi$, where $\sqrt{\gamma_2}$ is the
unit volume on $S^2$. Here $\vec x_2$ are the projections of $\hat n$ onto $S^2$. 
Since $\de_\eta$ is a global Killing vector of (\ref{metricsE}), 
the conserved inner product on the space of mode functions on $S^2$ is
\be 
 (\Psi, \Phi) = i \rho^2 \int  d{\cal V} \Bigl( \Psi^*(x) \, \overset\leftrightarrow{\partial_\eta} \, \Phi(x) \Bigr) \, ,
 \label{innerp}
\ee
where $d{\cal V} = d^2x \sqrt{\gamma_2}$ is the invariant measure 
on $S^2$, $\gamma_{2ab}$ metric on the unit sphere $S^2$, and 
$\Psi^*\, \overset\leftrightarrow{\partial_\eta} \, \Phi = \Psi^* ({\partial_\eta} \Phi) - ({\partial_\eta} \Psi^*) \Phi$. 
This inner product is conserved; using Green's formula for the field equation
(\ref{fullfeq}), for any two different solutions $\Psi, \Phi$ of (\ref{fullfeq}) we find that the 
``charge" deposited on a spacelike surface $\eta = {\rm const}$
\be
Q_{\Psi, \Phi} =  i \int  d^2x \sqrt{g_3} \, g^{0\,\nu} \Bigl( \Psi^*(x) \, \overset\leftrightarrow{\partial_\nu} \, \Phi(x) \Bigr) \, ,
\label{currentcons}
\ee
is conserved in time, since by (\ref{fullfeq}), 
$ \partial_\mu \Bigl(\sqrt{g_3} \, g^{\mu\nu} \bigl( \Psi^*(x) \, \overset\leftrightarrow{\partial_\nu} \, \Phi(x) \bigr) \Bigr) = 0$. 
On the static Einstein metric (\ref{metricsE}), equation (\ref{currentcons}) reduces to  (\ref{innerp}). 

Note that the conserved charge $Q_{\Psi, \Phi}$ defined in Eq. (\ref{currentcons}) is conformally invariant - i.e., it does not change 
under the transformations (\ref{weyl}). That is easy to see since (\ref{currentcons}) is scale invariant -- i.e. invariant under the rigid
rescalings (\ref{weyl}) with $\Omega = {\rm const}$. Those are lifted to the full local transformation invariance involving 
$\Omega(x)$ because $\Omega$ are real, and the charge (\ref{currentcons}) comes from the Klein-Gordon current whose
charge is zero for real fields. In more detail, Green's formula yields $\bar \Psi^*(x) \, \overset\leftrightarrow{\partial_\nu} \, \bar \Phi(x) = \frac{\Psi^*(x)}{\sqrt{\Omega}} \, 
\overset\leftrightarrow{\partial_\nu} \,  \frac{\Phi(x)}{\sqrt{\Omega}} = \frac{1}{\Omega}
\Psi^*(x) \, \overset\leftrightarrow{\partial_\nu} \, \Phi(x) $. 

The equal time commutation relations (\ref{eqticommrel}) then imply the standard dual domain commutation relations 
\be
[ a_{\ell m}, a_{\ell' m'}^\dagger] = \delta_{\ell \ell'} \delta_{m m'} \, , ~~~~~~~~
[ a_{\ell m}, a_{\ell' m'}]  = [ a_{\ell m}^\dagger, a_{\ell' m'}^\dagger] = 0 \, , 
\label{dualcommrel}
\ee
after imposing the normalization $(v_{\ell m}, v_{\ell' m'}) = \delta_{\ell \ell'} \delta_{m m'}$, which requires 
$|{\cal A}_{\ell m}|^2 = \frac{1}{(2 \ell + 1) \rho} = \frac{1}{2\om_\ell \rho^2}$. 
The complete set of orthonormalized eigenfunctions and their complex conjugates are
\be
v_{\ell m}(x^\mu) = \frac{1}{\sqrt{2 \om_\ell}\, \rho} Y_{\ell m}(\hat n) e^{- i \om_{\ell} \eta} \, , ~~~~~~~~
v^*_{\ell m}(x^\mu) = \frac{1}{\sqrt{2 \om_\ell} \, \rho} Y^*_{\ell m}(\hat n) e^{i \om_{\ell} \eta} \, .
\label{allmodes}
\ee
The time-dependence of the eigenmodes enters via an overall phase factor, which implies that there 
is no mode mixing induced by the free field evolution. Modes with a given eigenvalue evolve in time 
by shifting their phase, and retaining the same frequency. In particular, the number operator 
${\hat N}_{\ell m} = a^\dagger_{\ell m} a_{\ell m}$ is time-independent, since the phase factors cancel out. 
This means, that the unique eigenmode of all ${\hat N}_{\ell m}$ with zero eigenvalues, 
${\hat N}_{\ell m} | 0 \rangle_0 = 0$ behaves as a true vacuum state. Once selected
as the initial condition, it will never get spontaneously excited by time evolution alone. Other states
in the theory's Hilbert space are then constructed by raising operators $a_{\ell m}^\dagger$ acting on $|0\rangle_0$. 

Indeed, to proceed with the construction of the Hilbert space, let us first write out the
Hamiltonian of the quadratic theory. From (\ref{action}), when $g_6 = 0$,
\be
\hat {\cal H}_0= \frac{\rho^2}{2} \int d{\cal V} \Bigl(\frac{\pi^2}{\rho^4 \gamma_2} + \frac{1}{\rho^2} \gamma_2^{ab}\nabla_{2a} \phi
\nabla_{2b} \phi + \frac{1}{4\rho^2} \phi^2 \Bigr) \, ,
\label{freeham}
\ee
where $\nabla_2$ is the gradient operator on $S^2$. Using 
$\pi = \rho^2 \sqrt{\gamma_2} \dot \phi$, substituting (\ref{fop}) for $\phi$, and using 
$ \int d{\cal V} \gamma_2^{ab}\nabla_{2a} \phi
\nabla_{2b} \phi = - \int d{\cal V} \phi \nabla_\Omega^2 \phi$ since $S^2$ has no boundary, and 
remembering orthogonality of $Y_{lm}$'s, after a
straightforward algebra we find that the ``mass terms" are absorbed into frequencies
due to the conformal symmetry of the theory and the isometries of static Einstein, using
$\int d{\cal V} \, Y_{\ell m}(\hat n) Y^*_{\ell' m'}(\hat n') = \delta_{\ell \ell'} \delta_{mm'}$,
\be
\int d{\cal V} (- \phi \nabla_\Omega^2 \phi + \frac{1}{4\rho^2} \phi^2) 
= \frac14 \sum_{\ell m \ell' m'} \omega_\ell 
\Bigl( a_{\ell m}a^\dagger_{\ell' m'} + a^\dagger_{\ell m}a_{\ell' m'} \Bigr) \delta_{\ell \ell'} \delta_{mm'} \, ,
\ee
and finally,
\be
\hat {\cal H}_0= \frac12 \sum_{\ell m} \om_{\ell} \Bigl(a_{\ell m} a^\dagger_{\ell m} + a^\dagger_{\ell m} a_{\ell m} \Bigr) 
= \sum_{\ell m} \om_{\ell} \Bigl(\hat N_{\ell m} + \frac12 \Bigr) \, .
\label{freehamdual}
\ee
The zero-point energy term in (\ref{freehamdual}), $\propto \frac12 \sum_{\ell m} \om_{\ell m}$ is formally divergent 
without a hard cutoff; we remove it with normal ordering, as is common. This leaves us with the quadratic 
Hamiltonian
\be
\hat {\cal H}_0 = \sum_{\ell m} \om_{\ell} \, \hat N_{\ell m} \, ,
\label{freehamdualfin}
\ee
which describes the system of countably many decoupled harmonic oscillators on $S^2$. 

Since $[\hat N_{\ell m}, \hat N_{\ell' m'}] = 0$ and $\hat N_{\ell m}^\dagger = \hat N_{\ell m}$, we can simultaneously diagonalize all
$\hat N_{\ell m}$ operators (which obviously diagonalizes the Hamiltonian, too). Further, since 
\be
[\hat N_{\ell' m'}, a_{\ell m}] = - a_{\ell m} \, \delta_{\ell \ell'} \delta_{mm'} \, , ~~~~~~~~ [\hat N_{\ell' m'}, a_{\ell m}^\dagger] 
= a_{\ell m}^\dagger \, \delta_{\ell \ell'} \delta_{mm'} \, ,
\ee
then, if $| \{\lambda\} \rangle$ is an eigenstate of all $\hat N_{\ell m}$'s, 
$\hat N_{\ell m} | \{\lambda\} \rangle = \lambda_{\ell m} | \{\lambda\} \rangle$, we have
\ba
\hat N_{\ell m} a_{\ell' m'} | \{\lambda\} \rangle 
&=& \bigl(\lambda_{\ell m} - \delta_{\ell \ell'} \delta_{mm'} \bigr) a_{\ell' m'}| \{\lambda\} \rangle \, , 
\nonumber \\ 
\hat N_{\ell m} a^\dagger_{\ell' m'} | \{\lambda\} \rangle 
&=& \bigl(\lambda_{\ell m} + \delta_{\ell \ell'} \delta_{mm'} \bigr) a^\dagger_{\ell' m'}| \{\lambda\} \rangle \, , 
\label{numberalg}
\ea
implying that if $|\{\lambda \} \rangle$ is an eigenstate with eigenvalues $\lambda_{\ell m}$, then $a_{\ell' m'}|\{\lambda \} \rangle$  is an
eigenstate where the eigenvalue $\lambda_{\ell' m'}$ is reduced by one unit, and 
$a^\dagger_{\ell' m'}|\{\lambda \} \rangle$  is an
eigenstate where the eigenvalue $\lambda_{\ell' m'}$ is increased by one unit. 
If any $\lambda_{\ell m}$ were not an integer, the
sequence of states $| \{\lambda\}\rangle$ would be an unbounded sequence from both ends, and that would imply that the
Hamiltonian $\hat {\cal H}_0$ is unbounded. Such a system would be quantum-mechanically unstable. To ensure this is avoided,
therefore, we must take all $\lambda_{\ell m}$ to be integers, meaning that for any state $| \{ \lambda \}\rangle$ for
any $a_{\ell m}$ there exists an integer ${\cal N}_{\ell m}$ such that the eigenvalue $\lambda_{\ell m}^{({\cal N}_{\ell m})}$ 
of $a^{{\cal N}_{\ell m}}_{\ell m} | \{\lambda\}\rangle$ is zero. 

Hence as a corollary, if we take any state $ | \{\lambda\}\rangle$, after repeated
but finite number of actions of all the lowering operators $a_{\ell m} $ on it we will reach the state $| 0 \rangle_0$ such that
\be
\hat N_{\ell m} | 0 \rangle_0 = 0 \, ,  ~~~~~~~~ \hat {\cal H}_0 | 0 \rangle_0 = 0 \, .
\label{vacdef}
\ee
This state is the vacuum of the theory. Being non-degenerate, it is unique. 
Since the quadratic Hamiltonian depends only on $\hat N_{\ell m}$, we have 
$\partial_\eta | 0 \rangle_0 = 0$, or $e^{-i \eta \hat {\cal H}_0} | 0 \rangle_0 = | 0 \rangle_0$. 

Using the second of Eqs. (\ref{numberalg}), we can check that any other state in the theory can be obtained
from the action of $a^\dagger_{\ell m}$ on $| 0 \rangle_0$. Indeed, the `excited' states on top of the vacuum
$| 0 \rangle_0$ and their eigenvalues are
\ba
| \{{\cal N}_{\ell m} \}\rangle_0 &=& 
\prod_{\ell m} \frac{ \bigl(a^\dagger_{\ell m}\bigr)^{{\cal N}_{\ell m}}}{\sqrt{{\cal N}_{\ell m}! }}  | 0 \rangle_0 \, ,
 ~~~~~~~~~ 
\hat N_{\ell m} |  \{{\cal N}_{\ell m} \}\rangle_0 = {\cal N}_{\ell m}  |  \{{\cal N}_{\ell m} \}\rangle_0 \, , \nonumber \\
\hat {\cal H}_0 |  \{{\cal N}_{\ell m} \}\rangle_0 &=& {\cal E}_{ \{{\cal N}_{\ell m}\}}  |  \{{\cal N}_{\ell m} \}\rangle_0 \, , 
~~~~~~~~~~~~~~~~~~~  \,
{\cal E}_{ \{{\cal N}_{\ell m}\}} = \sum_{\ell m} \om_\ell \, {\cal N}_{\ell m}  \, .
\label{excited}
\ea

The states $| \{{\cal N}_{\ell m}\}\rangle_0$ provide a complete local basis of the Hilbert space of the quadratic theory at some time $\eta$,
and are taxonomized by the ``occupation numbers" ${\cal N}_{\ell m}$. Since they
are eigenstates of the quadratic Hamiltonian $\hat {\cal H}_0$, their local their time
evolution is obviously just given by a phase shift,
\be
| \{{\cal N}_{\ell m}\}(\bar \eta)\rangle_0 = e^{-i {\cal E}_{ \{{\cal N}_{\ell m}\}} (\bar \eta - \eta)} | \{{\cal N}_{\ell m}\}(\eta)\rangle_0 \, ,
\ee
which follows from the Schr\"odinger equation 
$i \partial_\eta | \{{\cal N}_{\ell m}\}(\eta)\rangle_0 = \hat {\cal H}_0 | \{{\cal N}_{\ell m}\}(\eta)\rangle_0$ ($\hbar=1$). 
Since time evolution is a phase shift, there is no Bogoliubov mode rotation by time evolution in {\it any} initial state. 
Thus there is no particle production, and in particular the definition
of the vacuum as the unique lowest energy state of the scalar on static Einstein is globally valid. As we noted, this is in fact true
for excited states as well. Once picked, the mode functions $v_{\ell m}, v_{\ell m}^*$ in (\ref{allmodes}) are preserved by time evolution, and so are the states $| \{{\cal N}_{\ell m} \}\rangle_0$, in the absence of interactions.

\begin{figure}[thb]
    \centering
    \includegraphics[width=9cm]{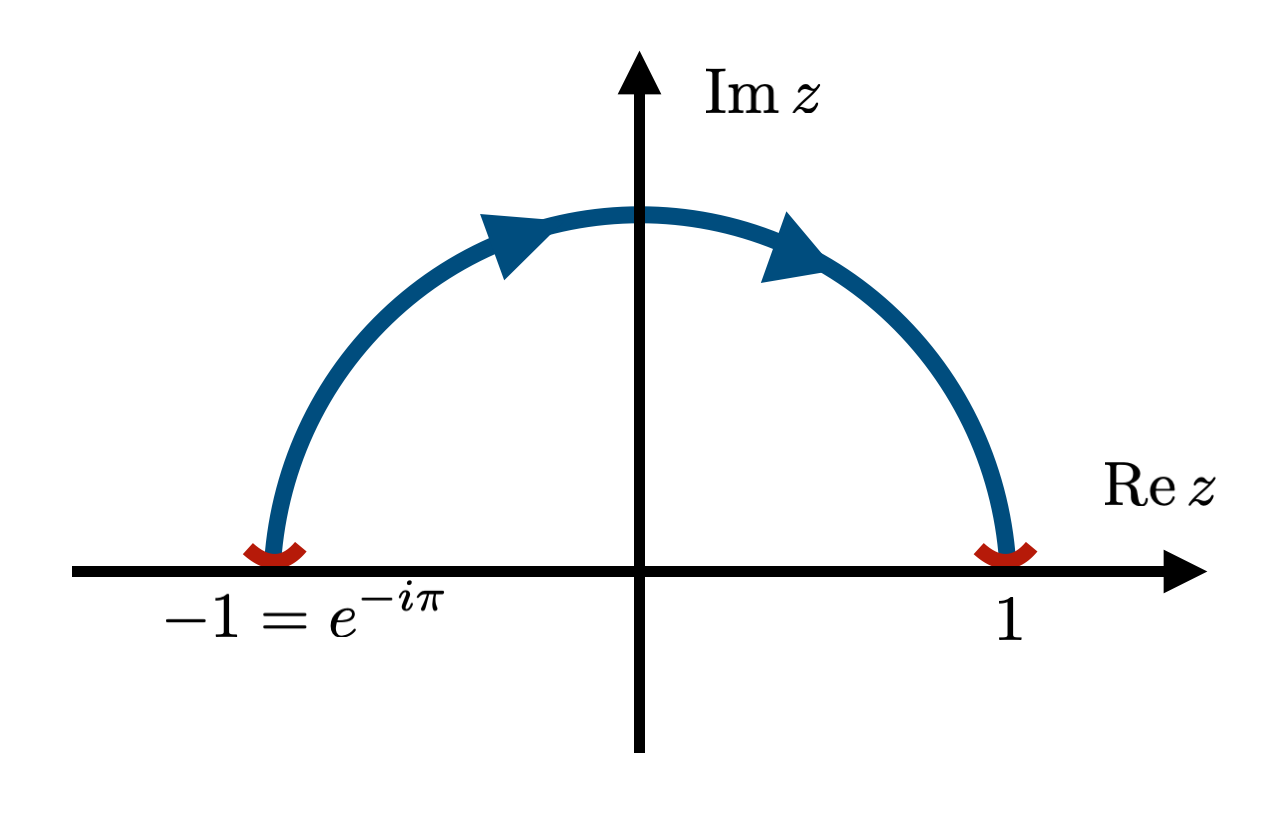} 
     \caption{de Sitter conformal gauge wavefunction phase time evolution.}
    \label{figsemi}
\end{figure}

It is interesting to see how the time evolution of these states looks like in the global coordinates chart of de Sitter. 
We can readily transform the wavefunctions (\ref{allmodes}) to the de Sitter coordinates by inverting the 
transformation (\ref{weyl}) with $\Omega = -\sin(\frac{\eta}{\rho})$ of Eq. (\ref{weyltrafo}) and substituting the 
conformal time $\eta$ with de Sitter global time $t$. The coordinate transformation is straightforward to find
by noting that (\ref{comconftime}) implies that $-\sin(\eta/\rho) = 1/\cosh(t/\rho)$, and so 
$\cos(\eta/\rho) = \tanh(t/\rho)$. Summing the two yields
\be
e^{-i\eta/\rho} = \frac{\sinh(t/\rho) + i}{\cosh(t/\rho)} = \frac{e^{t/\rho}+i}{e^{t/\rho}-i} \, .
\label{dsphases}
\ee
It is obvious that $|\frac{\sinh(t/\rho) + i}{\cosh(t/\rho)}| = 1$, and so the formula (\ref{dsphases}) is a map of the
real axis $t \in (-\infty,\infty)$ to the open unit semi-circle in the upper 
half-plane of the complex time variable, centered 
at the origin, as depicted in Fig. (\ref{figsemi}).
The wavefunctions (\ref{allmodes}) map to 
\ba
v_{\ell m}(x^\mu) &=& \frac{1}{\sqrt{2 \om_\ell}\, \sqrt{\cosh(t/\rho)} \, \rho} Y_{\ell m}(\hat n) 
\Bigl(\frac{\sinh(t/\rho) + i}{\cosh(t/\rho)}\Bigr)^{-\omega_\ell \rho} \, , 
\nonumber \\
v^*_{\ell m}(x^\mu) &=& \frac{1}{\sqrt{2 \om_\ell}\, \sqrt{\cosh(t/\rho)} \, \rho} Y^*_{\ell m}(\hat n) 
\Bigl(\frac{\sinh(t/\rho) - i}{\cosh(t/\rho)}\Bigr)^{\omega_\ell \rho} 
\, .
\label{allmodesds}
\ea
The time evolution in de Sitter `gauge' arises due to two effects. One is 
the wavefunction phase shift, which involves frequencies that change as de Sitter global time varies. 
In terms of the time $t$, a system point traverses the semi-circle in
Fig. (\ref{figsemi}) at a time-dependent rate, given precisely by the conformal 
transformation factor $\Omega$, since
$d\eta = - \sin(\eta/\rho) dt$, as specified in Eq. (\ref{comconftime}), instead of with a constant rate as in
variable $\eta$.

The other effect arises because of the normalization
factor $\sim 1/\sqrt{\cosh(t/\rho)}$ in (\ref{allmodesds}). The amplitude of the wavefunctions decreases to zero as $t \rightarrow \pm \infty$,
but at the rate which guarantees that the norm is preserved, so that the integral (\ref{currentcons}) remains constant. This dilution of the
wavefunctions can be viewed as the cosmological redshift effect in de Sitter.
Again, with this choice of Hilbert space basis both the ground state and all the excited states
remain unaffected by spontaneous mode creation induced by time evolution alone. 

\section{Revivals}

Extending time evolution beyond the interval $(-\pi \rho, 0^-)$ on $R \times S^2$
reveals interesting features of the topology of time axis. Let us consider the field operator $\phi$ 
given by (\ref{fop}), (\ref{allmodes}). Upon a time translation $\eta \rightarrow \eta + \eta_0$, we see from (\ref{allmodes}) that
\be
v_{\ell m}(x^\mu) \rightarrow v_{\ell m}(x^\mu) e^{- i \om_{\ell} \eta_0} \, , ~~~~~~~ 
v^*_{\ell m}(x^\mu) \rightarrow v^*_{\ell m}(x^\mu) e^{i \om_{\ell} \eta_0} \, .
\label{tphase}
\ee
This immediately shows that for $\eta_0 = 4\pi \rho$, the eigenfunctions $v_{\ell m}(x^\mu), v^*_{\ell m}(x^\mu)$, and so the field
$\phi$ and any state $|\{{\cal N}_{\ell m}\}\rangle_0$, complete a full cycle, e.g. 
\be
\phi(\eta+4\pi \rho) = \phi(\eta) \, ,
\label{perodicity}
\ee
because from (\ref{eigenv}), it follows that $\om_{\ell} \eta_0 = (2\ell +	1)2\pi$ 
and so the overall phase factor for both sets of modes is just unity, 
\be
e^{\mp i \om_{\ell} \eta_0} = e^{\mp i (2\ell +	1)2\pi}=1 \, , 
\label{phs1}
\ee
because $\ell$ are integers. This implies that there is no non-trivial quantum evolution beyond a conformal time interval 
$\Delta \eta = 4\pi\rho$ on $R \times S^2$. Once an initial condition is chosen, the evolution of $\phi$ {\it exactly} repeats 
after a time interval of length $4\pi \rho$. 

However, this is an overkill; the theory has additional redundancies that yield evolutionary repetition 
even without waiting for $4\pi \rho$. To see this, consider a time shift by $\eta_0 = 2\pi \rho$. In this case,
the overall phase factor in (\ref{tphase}) is  
\be
e^{\mp i \om_{\ell} \eta_0} = e^{\mp i (2\ell +	1)\pi}=-1 \, , 
\label{phs2}
\ee
because $2\ell+1$ are odd integers. 
This means that for $\phi$, the only change is the multiplicative factor of $-1$: 
\be
\phi(\eta+2\pi \rho) = - \phi(\eta) \, . 
\label{perodicity2}
\ee
Hence the dual domain operators $a_{\ell m}, a_{\ell m}^\dagger$ computed at a time $\eta + 2\pi \rho$
all pick up a $-1$ factor too. As a result, the Hilbert space basis vectors $|\{{\cal N}_{\ell m}\}\rangle_0$ transform to
\be
|\{{\cal N}_{\ell m}\}\rangle_0 \rightarrow \prod (-1)^{{\cal N}_{\ell m}} |\{{\cal N}_{\ell m}\}\rangle_0 \, .
\ee
Therefore the dot products of different state vectors remain unchanged; the only nonzero dot products must be of the form
\be
~_0\langle \{\bar {\cal N}_{\ell' m'}\}|\{{\cal N}_{\ell m}\}\rangle_0 \rightarrow \prod (-1)^{{\cal N}_{\ell m}} 
(-1)^{{\cal N}_{\ell' m'}} ~_0\langle \{\bar{\cal N}_{\ell' m'}\}|\{{\cal N}_{\ell m}\}\rangle_0 \, ,
\ee
and they can only be nonzero if ${\cal N}_{\ell' m'} = {\cal N}_{\ell m}$. But in this case, the factors of $-1$ square up to unity.

In more general cases, when the states are linear combinations of different Hamiltonian eigenstates, the  only nonzero contributions 
to the dot products come from components with equal occupation numbers, and again $-1$'s square up to unity in each nonzero
contribution. 
For the same reason, operator matrix elements $~_0\langle \Psi | {\cal Q} |\Phi \rangle_0$ 
are the same at $\eta$ and $\eta +2\pi \rho$. In this case, $|\Psi \rangle_0, |\Phi \rangle_0$ may have different occupation
numbers, but $|\Psi \rangle_0, Q |\Phi \rangle_0$ must have at least some contributions that have the same occupation numbers
if they are to be non-zero. So the factors of $-1$ come in pairs, and always square up.
Thus the observations in the theory certainly 
repeat after a time interval of length $2\pi \rho$. 

The two examples above illustrate the effect of symmetries of $R \times S^2$ for the conformally coupled
field $\phi$ on time evolution. First, the spectrum is discrete, with energies $\propto \ell + 1/2$. Secondly, the
presence of symmetries and the fact that quantum states are ray representations, invariant under phase shifts, 
ensures that observables repeat after discrete time translations. 

However, we have not yet fully exploited all the redundancies of the theory. Both de Sitter and static Einstein, 
with a conformally coupled scalar field theory on them, also have parity symmetry. 
This is manifest from the properties of spherical harmonics 
$Y_{\ell m}(\hat n)$. Under parity ${\cal P}$, $\hat n \xrightarrow{\cal P} - \hat n$ 
(i.e., $(\theta, \chi) \xrightarrow{\cal P} (\pi - \theta, \pi+\chi)$), and so
\be
Y_{\ell m}(\hat n) \, \xrightarrow{\cal P} \, {\cal P}Y_{\ell m}(\hat n) = Y_{\ell m}(-\hat n) = (-1)^\ell \, Y_{\ell m}(\hat n) \, .
\label{parity} 
\ee
Consider now a time translation by $\eta_0 = \pi \rho$. After this time shift, 
the overall phase factor in (\ref{tphase}) is
\be
e^{\mp i \om_{\ell} \eta_0} = e^{\mp i (2\ell +	1)\pi/2}=i^{\mp 1} (-1)^\ell \, , 
\label{phs22}
\ee
and so the eigenfunctions transform to, using (\ref{parity}), 
\ba
v_{\ell m}(x^\mu) &\xrightarrow{\cal P}& - (-1)^\ell \, i \, v_{\ell m}(x^\mu)  = - i {\cal P} v_{\ell m}(x^\mu) \, , \nonumber \\
v^*_{\ell m}(x^\mu) &\xrightarrow{\cal P}& + (-1)^\ell \, i \, v_{\ell m}^*(x^\mu) 
= + i {\cal P} v^*_{\ell m}(x^\mu) = \Bigl(- i {\cal P} v_{\ell m}(x^\mu)\Bigr)^* \, .
\label{tphasepi}
\ea

Returning to the field operator $\phi$, and splitting it into positive and negative frequency pieces, $\phi_\pm$, we find 
that after a time translation by $\eta_0 = \pi \rho$, 
\ba
\phi(\eta + \pi \rho) &=& \phi_{+}(\eta + \pi \rho) +  \phi_{-}(\eta + \pi \rho) \nonumber \\
&=& -i {\cal P} \Bigl(\phi_+(\eta) - \phi_-(\eta) \Bigr)  \nonumber \\
&=& -i {\cal P} \phi_+(\eta) + \Bigl(-i {\cal P} \phi^*_-(\eta) \Bigr)^* \, .
\label{decomp}
\ea
As a consequence, the observables at $\eta+\pi \rho$ coincide with the observables at $\eta$. 
This follows from the fact that
the overlap integrals between positive and negative frequency eigenmodes are zero due to the conservation of angular momentum 
$(\ell,m)$ (i.e., ``momentum" on $S^2$), enforced by isomorphisms of $S^2$. 

To see this, consider two wavefunctions $\Phi = ~_0\langle 0 |\phi(x^\mu) |\Phi \rangle_0$, 
$\Psi = ~_0\langle 0 |\phi(x^\mu) |\Psi \rangle_0$ and their conserved overlap integral (\ref{innerp}). At a time
$\eta$, it is 
\ba
~_0\langle \Phi(\eta) | \Psi(\eta) \rangle_0 &=& i \rho^2 \int  d{\cal V} \Bigl( \Phi^* \, \overset\leftrightarrow{\partial_\eta} \, \Psi \Bigr) = 
 i \rho^2 \int  d{\cal V} \Bigl( \bigl(\Phi^*_+ + \Phi^*_-\bigr) \, \overset\leftrightarrow{\partial_\eta} \, \bigl(\Psi_+ + \Psi_-\bigr)  \Bigr) 
 \nonumber \\
 &=&  i \rho^2 \int  d{\cal V} \Bigl( \Phi^*_+ \, \overset\leftrightarrow{\partial_\eta} \, \Psi_+ + 
\Phi^*_- \, \overset\leftrightarrow{\partial_\eta} \, \Psi_- + 
\Phi^*_- \, \overset\leftrightarrow{\partial_\eta} \, \Psi_+ +  
\Phi^*_+ \, \overset\leftrightarrow{\partial_\eta} \, \Psi_- 
\Bigr) \nonumber  \\
&=&   i \rho^2 \int  d{\cal V} \Bigl( \Phi^*_+ \, \overset\leftrightarrow{\partial_\eta} \, \Psi_+ + 
\Phi^*_- \, \overset\leftrightarrow{\partial_\eta} \, \Psi_- \Bigr)  \, .
\label{overlapscons}
\ea
The last two terms in the middle line of (\ref{overlapscons}) vanish, and the last line follows, because
the positive and negative frequencies overlap integrals are zero. As a demonstration, consider 
explicitly single-occupancy states with only one $\ell, m$ excitation being occupied, such that 
\ba
~_0\langle \Phi_+(\eta) | \Psi_-(\eta) \rangle_0 &=&  i \rho^2 \int  d{\cal V} \, \Phi^*_+ \, \overset\leftrightarrow{\partial_\eta} \, \Psi_- 
=   i \rho^2 \sum_{\ell,m,\ell',m'} \hat c_{\ell m}^\Psi \, \hat c_{\ell' m'}^\Phi \int  d{\cal V}  \Bigl( v_{\ell' m'}^*(x) 
\, \overset\leftrightarrow{\partial_\eta} \, {v}^*_{\ell m}(x) \Bigr)  \nonumber \\
&=&- \rho^2 \sum_{\ell,m,\ell',m'} (\om_{\ell} - \om_{\ell'}) \, \hat c_{\ell m}^\Psi \, \hat c_{\ell' m'}^\Phi \, 
e^{i (\om_{\ell} + \om_{\ell'}) \eta} \int  d{\cal V}  \, 
Y_{\ell' m'}^*(\hat n) {Y}^*_{\ell m}(\hat n) \nonumber \\
&=&- \rho^2 \sum_{\ell,m,\ell',m'} (\om_{\ell} - \om_{\ell' }) \, \hat c_{\ell m}^\Psi \, \hat c_{\ell' m'}^\Phi \, 
e^{i (\om_{\ell} + \om_{\ell'}) \eta} \, \delta_{\ell \ell'} \delta_{-m,m'} = 0 \, ,
\label{overlapsconszero}
\ea
which vanishes thanks to the orthogonality of momentum eigenmodes $Y_{\ell m}$. Similarly, 
we can see that $~_0\langle \Phi_-(\eta) | \Psi_+(\eta) \rangle_0 = 0$. This remains true for higher occupation number states as well, as is easy to check.
All that changes is that $\Phi, \Psi$ involve products of single excitation wavefunctions, and  $~_0\langle \Phi(\eta) | \Psi(\eta) \rangle_0$ 
turns out to be a multiple 
integral over $S^2$. 

At the time $\eta + \pi \rho$, the same is true. Using (\ref{decomp}),
$\Psi(\eta + \pi \rho) = -i {\cal P} \Bigl(\Psi_+(\eta) - \Psi_-(\eta) \Bigr)$, the overlap integral is, due to Hermiticity and parity 
idempotency, 
\ba
~_0\langle \Phi(\eta+\pi \rho) | \Psi(\eta+\pi \rho) \rangle_0 &=& 
 i \rho^2 \int  d{\cal V} \Bigl( i {\cal P} \bigl(\Phi^*_+ 
 - \Phi^*_-\bigr) \, \overset\leftrightarrow{\partial_\eta} \, (-i{\cal P})\bigl(\Psi_+ - \Psi_-\bigr)  \Bigr) 
 \nonumber \\
 &=&  i \rho^2 \int  d{\cal V} \Bigl( \Phi^*_+ \, \overset\leftrightarrow{\partial_\eta} \, \Psi_+ + 
\Phi^*_- \, \overset\leftrightarrow{\partial_\eta} \, \Psi_- - 
\Phi^*_- \, \overset\leftrightarrow{\partial_\eta} \, \Psi_+ -  
\Phi^*_+ \, \overset\leftrightarrow{\partial_\eta} \, \Psi_- 
\Bigr) \nonumber \\
&=&   i \rho^2 \int  d{\cal V} \Bigl( \Phi^*_+ \, \overset\leftrightarrow{\partial_\eta} \, \Psi_+ + 
\Phi^*_- \, \overset\leftrightarrow{\partial_\eta} \, \Psi_- \Bigr)  \, , 
\label{overlapsconsshift}
\ea
because the overlaps $~_0\langle \Phi_+(\eta) | \Psi_-(\eta) \rangle_0$ and $~_0\langle \Phi_-(\eta) | \Psi_+(\eta) \rangle_0$
still vanish by virtue of (\ref{overlapsconszero}), regardless of the algebraic manipulations which result from time
translation. Hence, as claimed, 
\be
~_0\langle \Phi(\eta+\pi \rho) | \Psi(\eta+\pi \rho) \rangle_0 = ~_0\langle \Phi(\eta) | \Psi(\eta) \rangle_0 \, .
\ee
As a consequence, the matrix elements of any operator covariant under the full de Sitter symmetry will obey
\be
~_0\langle \Phi(\eta+\pi \rho) | \, {\cal Q} \, | \Psi(\eta+\pi \rho) \rangle_0 = ~_0\langle \Phi(\eta) | \, {\cal Q} \, | \Psi(\eta) \rangle_0 \, .
\ee

Therefore we conclude that the quantum evolution of {\it any} observable on 
static Einstein $R \times S^2$ is completely contained in any timelike interval of length $\Delta \eta = \pi \rho$.
Going beyond any such interval simply yields repetitions of the ``already seen" physics for all quantum states. Once we pick 
a spacelike surface in $R \times S^2$ at a fixed time $\eta$, we only need to move the states defined on it a finite
amount of $\Delta \eta = \pi \rho$ to collect all the information about the system. In the absence of a UV cutoff, this
still involves infinitely many eigenstates $v_{\ell m}$, but as they are discrete the Hilbert space census only involves counting. 

Since time evolution of observables is periodic, and the period is completely independent of the 
particular state the system is in, we infer that the time axis is, for all practical intents and purposes, 
effectively restricted to an interval of width 
$\Delta \eta = \pi \rho$, for a conformal gauge-fixing of the radius of $S^2$ to $\rho$. Beyond this interval,
the wavefunction evolution repeats the already recorded past history from any specification of the 
initial state, where the evolution in the adjacent interval goes ``backwards", as we noted above. 
Since the nontrivial pattern of any wavefunction is completely described 
in this compact time interval, we can formally restrict 
$R \times S^2$ to $I \times S^2$. This does not mean we are imposing 
the past and future time boundaries by hand; we are merely observing the periodicity of quantum evolution of the system. 

After this time ``compactification", we recognize that the conformally coupled scalar theory on $R \times S^2$ is really the theory
on $dS_{2+1}$. At first sight this seems odd: in a fixed coordinate chart, $I \times S^2$ has fewer explicit 
isometries than de Sitter in $2+1$: the static Einstein segment $I \times S^2$ has 4 Killing vectors while $2+1$ de Sitter 
has 6 Killing vectors. However, there are two additional symmetries for a conformally coupled scalar on this background: 
the two special conformal transformations on $I \times S^2$. 

Hence we see that, once the symmetries are accounted for, a 
conformally coupled scalar on $2+1$ static Einstein is realization of the theory of a conformally coupled scalar in $2+1$ de Sitter, 
with manifest phenomenon of quantum revivals. We are able to go around the deep problems with de Sitter horizons by lifting 
the theory to $R \times S^2$, where there are no horizons, and formulate it there by finding the complete Hilbert space, 
while preserving covariance. The observables of the quantized theory feature full quantum revivals, which restrict the 
dynamics only to the nontrivial time intervals between two complete repetitions of the observables. This in turn  restricts the
theory back to de Sitter. 

Note that once we fix the conformal gauge to static Einstein, with the metric (\ref{metricsE}), we actually haven't 
completely fixed the gauge. There still is a residual scale transformation
\be
ds^2 \rightarrow d\tilde s^2 = \frac{\tilde \rho^2}{\rho^2} ds^2 = -d\tilde \eta^2 + \tilde \rho^2 d\Omega_2 \, , ~~~~~~~~~ 
d\eta \rightarrow d\tilde \eta = \frac{\tilde \rho}{\rho} \, d \eta \, .
\label{scaleres}
\ee
This transformation also changes the periodicity of the quantum wavefunction evolution,
\be
\Delta \eta \rightarrow \Delta \tilde \eta = \frac{\tilde \rho}{\rho} \, \Delta \eta = \pi \tilde \rho \, .
\label{scaleperiod}
\ee
Therefore the actual scale of the period $\rho$ is not really physically meaningful, since scaling transformations (\ref{scaleres})
can change it at will. What matters is that for any finite $\rho$, we have a countable basis of states that describe the
full theory, and that their complete and faithful evolution is completely inscribed in a finite time interval. There is nothing more to see
by waiting longer that hasn't been seen yet. In this way, the extension of the states beyond a single
period is trivial, but the construction gets around the horizons 
which would limit the access to such information in de Sitter. 

\section{Blowing up the Sphere: In\"on\"u-Wigner} 

\begin{figure}[thb]
    \centering
    \includegraphics[width=11cm]{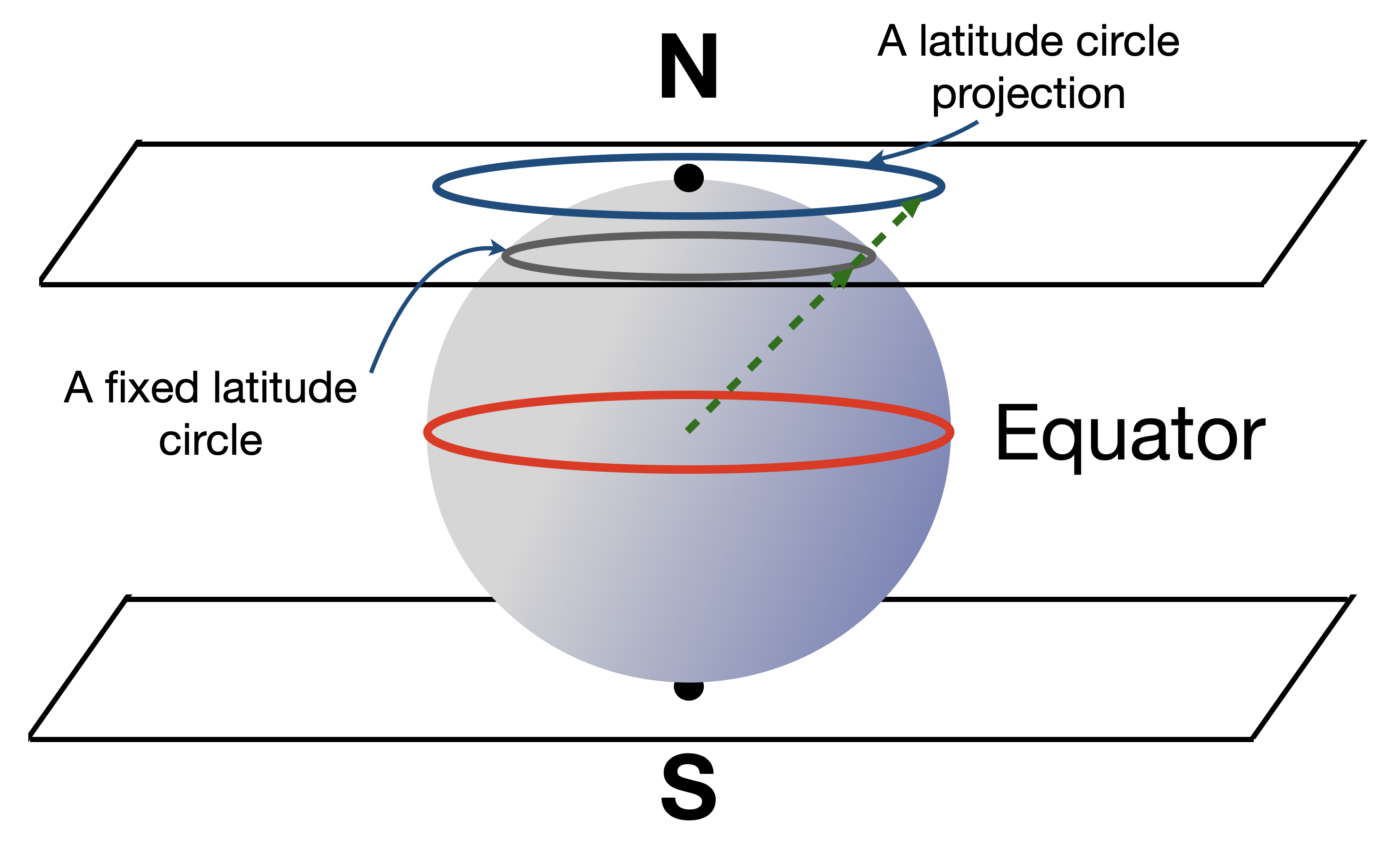} 
     \caption{Blowing up the $S^2$: stereographic projections from the center of the sphere.}
    \label{fig3}
\end{figure}

It is interesting to see what happens in the limit $\rho \rightarrow \infty$. In this case, the radius of the 
sphere becomes infinite, and we should take it with some care. The procedure is illustrated in Fig. (\ref{fig3}). 
Basically, what happens is that we take polar caps at a fixed radius $\rho$ and blow them up huge, by effectively making the radius 
of the equator infinite. This leaves us with two copies of the infinite hemisphere -- the two disconnected 
copies, each a half of $S^2$, anchored at North and South pole. 

Mathematically, we can perform this limit by coordinatizing static Einstein metric (\ref{metricsE}) 
using the constrained Cartesian coordinates ${x,y,z}$ with $x^2 + y^2 + z^2 = \rho^2$, eliminating $z$ and introducing $2D$
polar spatial coordinates $r, \chi$ by $x = r \cos \chi, y = r \sin \chi$. Then, after a straightforward algebra, 
the metric (\ref{metricsE}) becomes 
\be
ds^2 = - d\eta^2 + \frac{\rho^2}{\rho^2 - r^2} dr^2 + r^2 d\chi^2 \, .
\label{rchimetric}
\ee
For a fixed $\rho$, using $r = \rho \sin\theta$ recovers (\ref{metricsE}). On the other hand, taking the limit 
$\rho \rightarrow \infty$ reproduces $2+1$ flat space -- two copies of it, actually, once we populate the $S^2$ with the
scalar theory which solves (\ref{freefeq}). In this limit, (\ref{freefeq}) simplifies to
\be
 - \de_\eta^2 \phi + \nabla_{\vec x}^2 \phi = 0 \, ,
 \label{freefeqflat}
\ee
the energy eigenvalues of (\ref{eigenv}) become a continuous variable $\om$, 
since $\om_{\ell} = \frac{2\ell +	1}{2\rho}$ implies that $\Delta \om = 1/\rho \rightarrow 0$, and so 
(\ref{freefeqflat}) in polar coordinates, where $\nabla_{\vec x}^2 \phi = \partial_r^2 \psi + \partial_r \psi/r + \partial_\chi^2 \psi/r^2$,
after separating variables $\psi = {\cal U} e^{\pm i m \chi}e^{\pm i \om \eta}$
becomes
\be
r^2 \partial_r^2 {\cal U} + r \partial_r {\cal U} + \Bigl( \om r^2 - m^2 \Bigr) {\cal U} = 0 \, .
 \label{freefeqflat1}
\ee
The radial modes are obviously Bessel functions, ${\cal U} = {\cal A}_\om J_m(\sqrt{\om} r)$, 
where ${\cal A}_\om$ is a normalization constant, and where we do not include Neumann's 
functions because they are singular at vanishing argument (which coordinatizes the poles).  
The mode functions in polar coordinates of the blown up hemispheres obtained by 
solving the theory (\ref{freefeqflat}) are therefore 
\be
\psi = {\cal A}_\om J_m(\sqrt{\om} r) e^{\pm i m \chi}e^{\pm i \om \eta} \, ,
\label{flat21modes}
\ee
and their complex conjugates.  
It should be clear that this procedure is In\"on\"u-Wigner contraction \cite{Inonu:1953sp} of the spatial sections of
static Einstein, since the spatial symmetry of $S^2$, which is $SO(3)$ is deformed to $ISO(2)$ when $r \rightarrow \infty$.
In principle, this means that the mode functions in the $\rho \rightarrow \infty$ limit can be obtained from (\ref{allmodes}), bearing
in mind that the spectrum becomes continuous, but we will leave this aside for now.

We note that the mode functions in this limit live on an infinite time interval since 
$\om$ is a continuous variable, and manifest periodicity in time, found when $\rho$ is finite, is now lost. The theory resides on a single 
$I \times S^2$ section of the covering space, which is blown up to be infinitely large in all directions. This is manifest 
from looking at $\Delta \eta = \pi \rho$ and taking the limit $\rho \rightarrow \infty$. 

If we were to go back to de Sitter global chart, and carry out $\rho \rightarrow \infty$ limit, our two planes obtained by 
blowing up the two hemispheres would in fact map onto two static patches of de Sitter in the limit when de Sitter radius goes to
infinity, and its curvature vanishes. Thus this limit can be viewed as the two disjoint static patches being infinitely blown up, 
and cooled to vanishing Gibbons-Hawking temperature. 

In this limit, the free conformally coupled scalar theory which solves (\ref{freefeqflat}) has uncountably many eigenmodes --
a lot more than there is on $R \times S^2$ with a fixed, finite sphere radius. These modes can be chosen to be (\ref{flat21modes}),
or massless plane waves in $2+1$ flat space. In fact, it may even seem simple to turn on the interactions: if we bring in
$\propto g_6 \phi^6$, the theory looks like a superrenormalizable QFT in $2+1$ Minkowski space, which is straightforward to formulate
since there are no horizons, and one might be tempted to use the standard asymptotic states and S-matrix definition in this limit.

As we warned earlier, if we change the conformal gauge of this theory back 
to $R \times S^2$ with a finite radius of $S^2$, there will be many fewer modes: the eigenmodes of the theory are 
gapped, and so countably infinite, as is clear from (\ref{allmodes}). This comes about due to the change of boundary conditions introduced by the In\"on\"u-Wigner contraction, which maps $S^2$ onto $R^2$, introducing a ``hole" 
on $S^2$ (i.e. treating equator as a boundary). It is interesting to explore this limit further. We will sidestep this question here, and instead 
focus on the theory on finite $S^2$, where the sparser spectrum and the boundary conditions on $S^2$ 
restrict time evolution, as we discussed above. 

\section{Sources}

\begin{figure}[thb]
\hskip0.3cm    \centering
    \includegraphics[width=7.5cm]{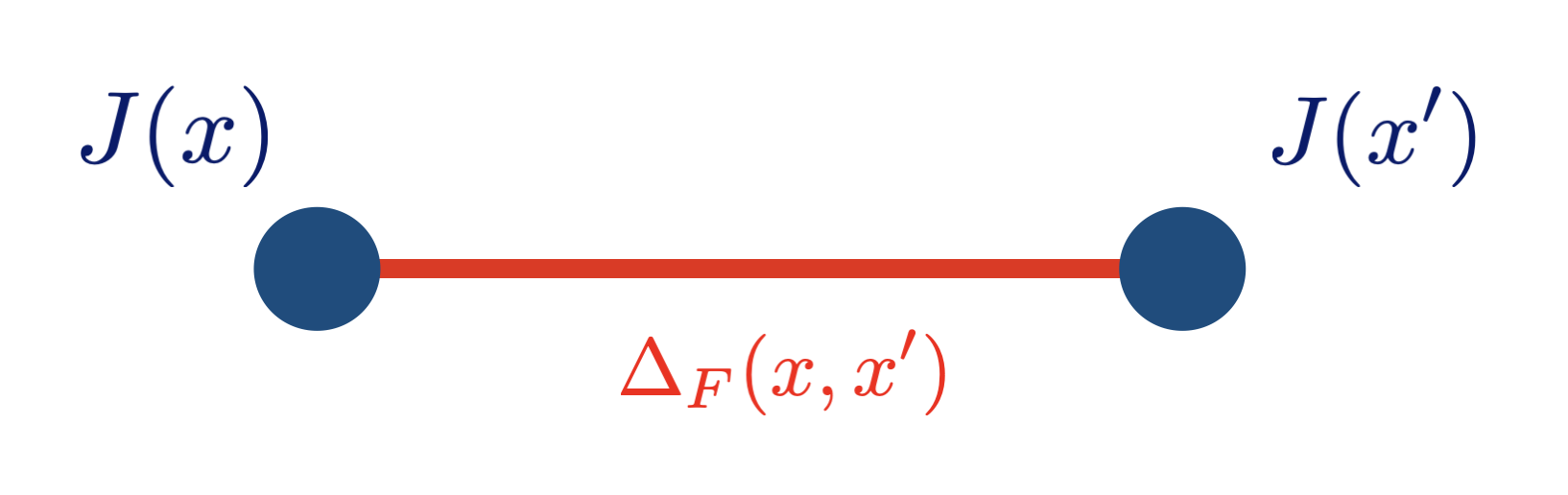} 
     \caption{J--J interaction.}
    \label{fig5}
\end{figure}

Because the combinatorial structure of the conformally coupled scalar field theory with sources on a fixed static Einstein background
is the same as it is in flat space, we can calculate the source interactions mediated by $\phi$ exchange straightforwardly. Diagrammatically,
this amounts to counting and summing up all the processes comprised of $1PI$ graphs of Fig. (\ref{fig5}). 
These diagrams exponentiate, as
can be readily seen from looking at the path integral for the quadratic theory of (\ref{action}) (with $g_6=0$)
on a fixed background (\ref{metricsE}): 
\be 
W(J) = \int \Bigl[{\cal D}\phi \Bigr] e^{i\int d^3x \sqrt{g}  \Bigl( \frac12 \phi \(\square_3  - \frac{1}{8} R \) \phi  +  \phi J \Bigr)} \, .
\label{path}
\ee
Here we integrated the scalar derivative terms
by parts to rewrite them in terms of the wave operator 
${\cal K} = \square_3 - \frac{1}{8} R$. On the static Einstein metric (\ref{metricsE}), the wave operator is
${\cal K} = -\partial_\eta^2 + \frac{1}{\rho^2} \nabla_\Om^2 - \frac{1}{4\rho^2}$. Next, as is common, we obtain the 
solution of the wave equation $\varphi$ with the source $J$, ${\cal K} \varphi = - J$, which can be written down by inverting 
${\cal K}$ with appropriately chosen boundary conditions:
\be
\varphi(x) = - {\cal K}^{-1} {\cal J}(x) = - \int d^3x' \sqrt{g} \, \Delta_F(x-x') J(x') \, ,
\label{wavesoln}
\ee
where $\Delta_F(x-x')$ is the Feynman propagator on $I \times S^2$, 
\be
{\cal K} \Delta_F(x-x') = \frac{1}{\sqrt{g}} \, \delta^{(3)}(x-x') \, .
\label{propagator}
\ee
The choice of the Feynman propagator $\Delta_F(x-x')$ is enforced by picking the boundary conditions 
for the Green's function, so as to ensure Dirac's interpretation that evolution forward
in time involves only particles, and evolution backward in time arises from antiparticles:
\ba
i \Delta_F(x-x') &=& ~_0\langle 0 | T\bigl(\phi(x) \phi(x') \bigr) | 0 \rangle_0 \nonumber \\
&=& \theta(\eta - \eta') 
~_0\langle 0 | \phi(x) \phi(x') | 0 \rangle_0 + \theta(\eta' - \eta) 
~_0\langle 0 | \phi(x') \phi(x) | 0 \rangle_0 \, .
\label{fprop}
\ea
Here, $T$ is the time-ordering operator and $\theta(x)$ the Heaviside step function. 

We can implement the boundary conditions using the 
$i \epsilon$ prescription for circumnavigating the momentum domain poles by replacing  
${\cal K} \rightarrow -\partial_\eta^2 + \frac{1}{\rho^2} \nabla_\Om^2 - \frac{1}{4\rho^2} + i \epsilon$, with our 
metric signature convention. This prescription reproduces the standard short distance behavior of flat space Green's function
in the limit $\rho \rightarrow \infty$. In the time domain, we enforce this 
boundary condition prescription by replacing $\eta - \eta' \rightarrow \eta - \eta' - i \epsilon$. 
The selection of positive versus negative frequency virtual excitation contributions to
momentum transfer then corresponds to the selection of how to close the integration contour in the complex time plane, controlled by
Jordan's lemma. In any case, this means that we can write down the propagator in terms of the Wightman functions 
$~_0\langle 0 | \phi(x) \phi(x') | 0 \rangle_0$. 

Then, shifting the path integral ``dummy" variable $\phi$ to
$\phi = \bar \phi + \varphi$ and substituting this into the path integral (\ref{path}) we obtain
\be 
W(J) = \int \Bigl[{\cal D}\bar \phi \Bigr] e^{i\int d^3x \sqrt{g}  \Bigl( \frac12 \bar \phi {\cal K} \bar \phi  - \frac12 J {\cal K}^{-1} J \Bigr)} \, ,
\label{path2}
\ee
where the path integral measure does not change, $\Bigl[{\cal D} \phi \Bigr] = \Bigl[{\cal D} \bar \phi \Bigr]$, since the Jacobian
induced by the field redefinition $\phi = \bar \phi + \varphi$ is unity. Since the integrand is a Gaussian in $\bar \phi$, the integration 
over $\bar \phi$ merely changes the normalization of $W$, and is trivially factored out. Thus, as is well known,
\be 
W(J) = e^{-i\int d^3x \sqrt{g} \frac12 J {\cal K}^{-1} J } = e^{-i\iint d^3x \sqrt{g(x)} \, d^3x' \sqrt{g(x')} \frac12 J(x) \Delta_F(x-x') J(x') } \, .
\label{path3}
\ee

Given the structure of the Hilbert space in Sec. (4), we can compute these objects straightforwardly. We have, using 
$~_0\langle 0 | a_{\ell m} a^{\dag}_{\ell' m'} | 0 \rangle_0 = \delta_{\ell \ell'} \delta_{m m'}$ and Eq. (\ref{sphaddth}), 
\ba
~_0\langle 0 | \phi(x) \phi(x') | 0 \rangle_0 = \sum_{\ell, m, \ell', m'} 
\frac{1}{\sqrt{4 \omega_\ell \omega_\ell' } \,\rho^2} e^{- i (\omega_\ell \eta- \omega_\ell' \eta')} 
Y_{\ell m}(\hat{n}) Y^*_{\ell' m'}(\hat{n}') \delta_{\ell \ell'} \delta_{m m'} ~~~~~~~~~~~~~~~~~~ && \nonumber\\
= \sum_{\ell, m} \frac{1}{2 \omega_\ell \, \rho^2} e^{- i \omega_\ell (\eta- \eta')} Y_{\ell m}(\hat{n}) Y^*_{\ell' m'}(\hat{n}')   
= \frac{1}{4 \pi \rho} \sum_{\ell} e^{- i \omega_\ell (\eta-\eta')} P_{\ell}(\hat{n} \cdot \hat{n}') \, .  &&  
\label{wightser}
\ea
This series can be summed to the generating function of Legendre polynomials, using the identity
\be 
  \frac{\sqrt{\zeta}}{\sqrt{1-2 \alpha \zeta +\zeta^2}} = \sum_{\ell=0}^\infty \zeta^{\ell+\frac{1}{2}} P_\ell(\alpha) \, .
  \label{sumPl}
\ee
Then, setting $\zeta = e^{-i \frac{\eta-\eta' -i \eps}{\rho}}$, $\alpha = \hat{n} \cdot \hat{n}'$, and recalling that 
$\omega_\ell = (\ell + 1/2)/\rho$, we recognize that it follows that
\be
~_0\langle 0 | \phi(x) \phi(x') | 0 \rangle_0  = \frac{1}{4 \sqrt{2} \pi \rho\sqrt{\cos\(\frac{\eta-\eta'-i \eps}{\rho}\) - \hat n \cdot \hat n'} }\, ,
\label{wighty}
\ee
where in the last step we restored the $i\epsilon$ prescription. 

The equations (\ref{wightser}) and (\ref{wighty}) provide further insight into the dynamics of $\phi$ on static Einstein.
First, it is evident that the momentum transfer between two sources, which is mediated by the Feynman propagators,
involves only the intermediate states which are fully described, channel by channel, by the nonlocal modes 
$P_\ell(\hat n \cdot \hat n')$. 

In other words, the local states $Y_{\ell m}(\hat n)$ always recombine into the many fewer
intermediate states $P_\ell$, enforced by the symmetries which constrain the dynamics. 
The modes $Y_{\ell m}$ are only necessary when we want to retain explicit information about
mutual relations of multiple observers on $S^2$: i.e. the dictionary allowing an observer (a) to immediately
translate the observations of observer (b) about an event at (c). But we can do this post-facto, by
replacing the results by 
$P_\ell(\hat n \cdot \hat n')$. 

Secondly, the Green's functions  (\ref{wightser}) and (\ref{wighty}) confirm the restriction of time evolution 
from $R \times S^2$ to $I \times S^2$ which we discussed in Sec. (5). Indeed, as there, we see that shifting 
$\eta \rightarrow \eta + \pi \rho$ in (\ref{wightser}) produces
\ba
~_0\langle 0 | \phi(x) \phi(x') | 0 \rangle_0 &\rightarrow& \frac{ -i }{4 \pi \rho} \sum_{\ell} e^{- i \omega_\ell (\eta-\eta')} 
(-1)^\ell P_{\ell}(\hat{n} \cdot \hat{n}') =  \frac{ -i }{4 \pi \rho} 
\sum_{\ell} e^{- i \omega_\ell (\eta-\eta')} P_{\ell}(- \hat{n} \cdot \hat{n}') ~~~ \nonumber \\
&=& \frac{ -i }{4 \pi \rho} \sum_{\ell} e^{- i \omega_\ell (\eta-\eta')} P_{\ell}(\hat{n} \cdot {\cal P} \hat{n}') \, , 
\label{wightsershift}
\ea
where ${\cal P}$ is the parity operator, 
implying that after a time interval $\pi \rho$, the evolution of the system reverses, beginning to trace back to the initial
configuration, starting from the antipode of each initial point. 

The resummed Green's function in Eq. (\ref{wighty}) shows this behavior directly too. The time 
shift $\eta \rightarrow \eta + \pi \rho$ yields, using 
$\cos\(\frac{\eta-\eta'-i \eps}{\rho} + \pi\) = -\cos\(\frac{\eta-\eta'-i \eps}{\rho}\)$, yields 
\be
~_0\langle 0 | \phi(x) \phi(x') | 0 \rangle_0 \rightarrow 
\frac{1}{4 \sqrt{2} \pi \rho\sqrt{-\cos\(\frac{\eta-\eta'-i \eps}{\rho}\) - \hat n \cdot \hat n'} } = 
\frac{-i}{4 \sqrt{2} \pi \rho\sqrt{\cos\(\frac{\eta-\eta'-i \eps}{\rho}\) - \hat n \cdot {\cal P} \hat n'} } \, .
\label{wightyshift}
\ee
This means, that the evolution of the system over $R \times S^2$ proceeds like the spirals on the barber pole: starting 
from one end of the pole, the evolution follows the null lines to the antipodal point, where the observables are the same 
as at the initial point. Beyond this, further evolution returns to the initial point, completing the full circle. Thus, there 
is no independent evolution beyond any $\Delta \eta = \pi \rho$ time interval. 

\section{Counting the Basis Dimension}

As noted above the conventional counting of field theory degrees of freedom in fact overcounts the 
modes which are necessary to store intrinsic dynamics. 
This occurs since a conformally coupled scalar on static Einstein $R \times S^2$ 
has a great degree of symmetry. Many configurations, virtual as well as real, on a fixed background will
be related to each other by symmetry.
\begin{figure}[thb]
    \centering
    \includegraphics[width=10.5cm]{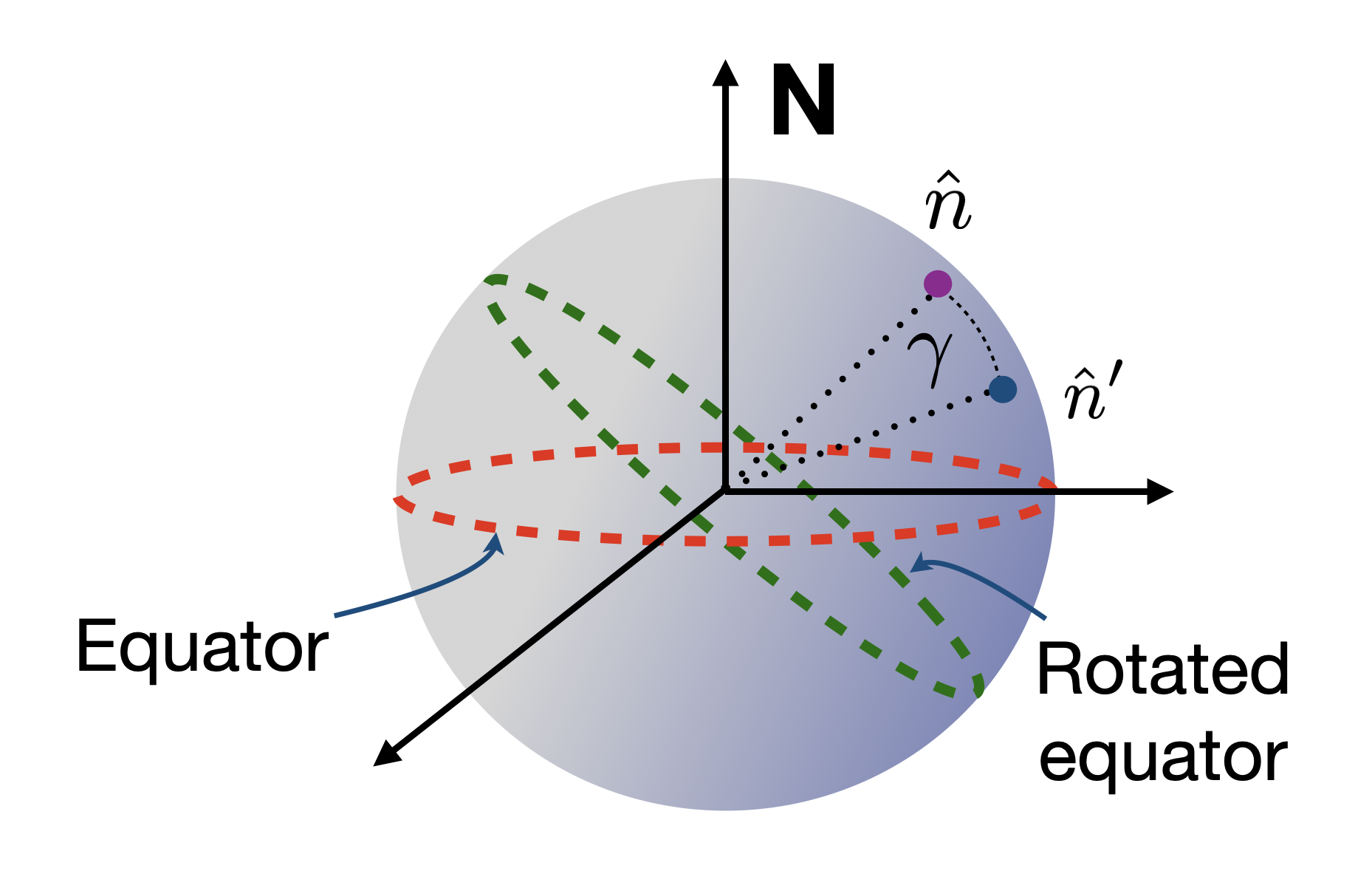} 
     \caption{Bases on $S^2$. Here $\gamma$ is the angle between two unit vectors
     $\hat n$ and $\hat n'$, such that $\cos(\gamma) = \hat n \cdot \hat n'$.}
    \label{fig4}
\end{figure}
We noted that this occurs for the theory on $R \times S^2$ with a finite radius of $S^2$, since the intrinsic dynamics 
can be encoded using fewer degrees of freedom than the basis of the local modes. 
The residual $S^2$ rotations restrict the form of the $2\ell+1$ eigenmodes 
$Y_{\ell m}(\hat n)$, which are subject to constraints such as the spherical 
harmonics addition theorem \cite{Bander:1965rz,Bander:1965im}
$$
P_\ell(\hat n \cdot \hat n') = \frac{4\pi}{2\ell+1} \sum_{m=-\ell}^\ell Y_{\ell m}(\hat n) Y^*_{\ell m}(\hat n') \, .
$$
The local basis $Y_{\ell m}$ does not take this into account directly. The issue is that, as we already 
noted in Sec. (2), we can represent any function on $S^2$ as a linear combination
of $P_\ell$ -- one for each $\ell$ -- instead of $2\ell+1$ functions $Y_{\ell m}$'s
$$
\Psi(\hat n) = \sum_{\ell,m} Y_{\ell m}(\hat n) \int_{S^2} d \hat n' \, Y^*_{\ell m}(\hat n') \Psi(\hat n') = 
\frac{2\ell+1}{4\pi} \sum_\ell \int_{S^2} d \hat n' \, P_\ell(\hat n \cdot \hat n') \Psi(\hat n') \, ,
$$
where, as noted previously, $d\hat n$ is the invariant measure on $S^2$. The price to pay is to replace $2\ell+1$ {\it local}
functions $Y_{\ell m}(\hat n)$ by a single nonlocal function  $P_\ell(\hat n \cdot \hat n')$.
Then, to restore locality we can, e.g., expand $P_\ell$ in its Taylor series around any 
``reference location" $\hat n$, by expanding the argument in
$\hat n' = \hat n + \delta n$. Since $P_\ell$'s are polynomials, 
the series truncates at finitely many terms, and the coefficients are precisely $Y_{\ell m}$. 

\begin{figure}[thb]
    \centering
    \includegraphics[width=13cm]{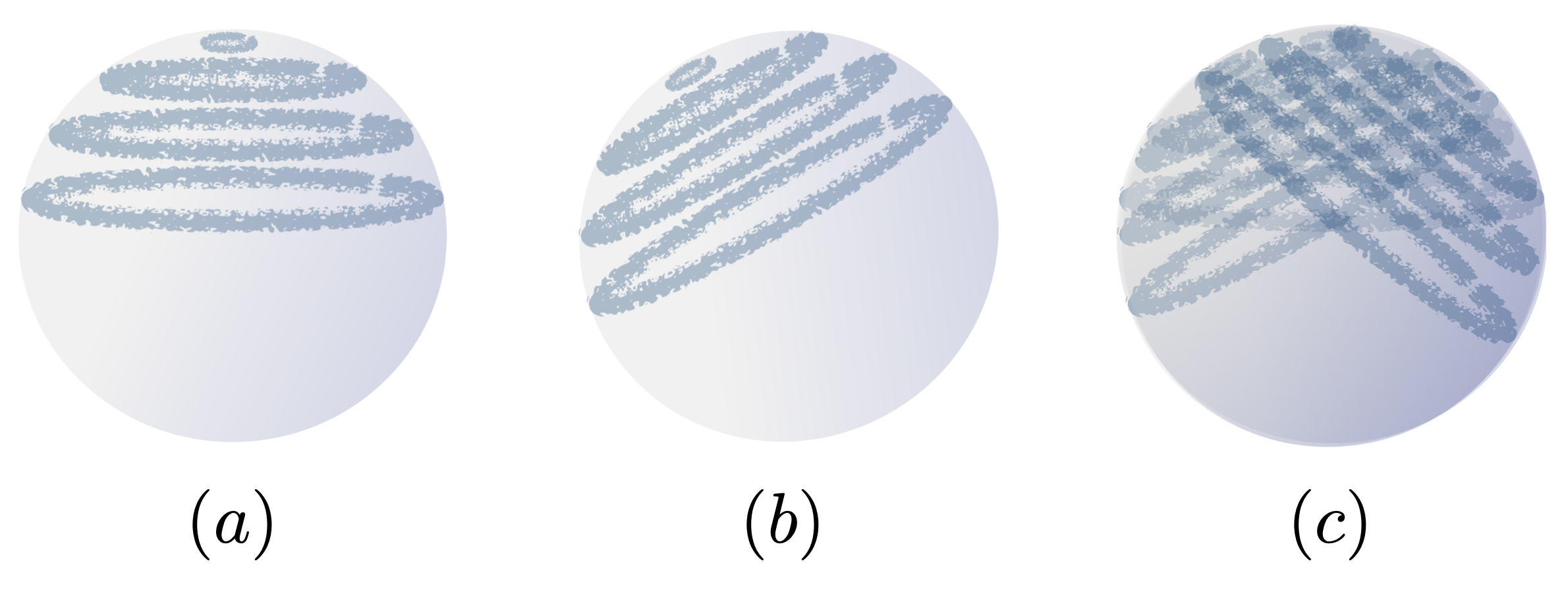} 
     \caption{Interference on $S^2$. We sketch $P_8$ in the Northern hemisphere 
     as a function of polar angle on $S^2$; panel (a) is with the symmetry axis
     along the vertical; panel (b) has the symmetry axis tilted by $30 ^{\circ}$ relative to the vertical; panel (c) is the superposition of 
     the three $P_8$, one symmetric about the vertical, one about an axis tilted by $30 ^{\circ}$ to the left and the final about an axis 
     titled by $40 ^{\circ}$ to the right. Dark spots are extinctions and bright spots amplifications. With sufficiently many ring systems we can 
     devise a profile with all but one points on $S^2$ extinguished.}
    \label{figint}
\end{figure}

This is illustrated
in Figs. (\ref{fig4}) and  (\ref{figint}). The information stored in the local basis $Y_{\ell m}$ in the coordinate system
anchored to the North Pole is completely equivalent to the information in the nonlocal basis spanned by fewer zonal
harmonics $P_\ell$, which depend on the nonlocal argument $\hat n \cdot \hat n' = \cos(\gamma)$. 

Indeed, the point of Fig. (\ref{fig4}) is to illustrate that when we consider the influence of a point source, placed somewhere 
on a sphere -- e.g., the North Pole -- the rotational symmetry of the theory about the $z$-axis states that the 
influence measured along any longitude line must be the same. However, when the source is displaced to a different
point, the local polar symmetry axis is tilted relative to the vertical, and we need the $Y_{\ell m}$'s to parameterize the
field of the source along the original longitudes. The apparent variation of the field 
at a fixed latitude as longitude is changed is merely induced by the tilt, since the rotational symmetry still restricts the form of the field 
to be invariant under rotations around the tilted polar axis $\hat n$. 

Further, the completeness
relation for local modes, $\sum_{\ell,m} Y_{\ell m}(\hat n) Y^*_{\ell m}(\hat n') = \delta(\hat n - \hat n')$ and the addition theorem imply 
that the nonlocal fucntions $P_\ell(\hat n \cdot \hat n')$ are also complete, 
\be
\sum_\ell \frac{2\ell+1}{4\pi} P_\ell(\hat n \cdot \hat n') =  \delta(\hat n - \hat n') \, .
\label{plcomplete}
\ee 
Hence, as depicted in Fig. (\ref{figint}), by linearly superposing sufficiently many $P_\ell$'s, 
we can recover localized lumps anywhere on $S^2$.

From this viewpoint, it is clear that the basis of the de Sitter Hilbert space may be significantly smaller than the naive counting of 
$Y_{\ell m}$ would suggest. Even so, if the theory is exactly conformal, even after we switch to the nonlocal 
basis of $P_\ell$'s, the Hilbert space is still discretely infinite. This implies that 
the counting of the basis must be UV-dominated, which is readily confirmed by 
the counting of the total number of states in the Hilbert space, which is also UV dominated, as is immediately 
seen from, e.g., the classic entanglement entropy 
counting in quantum theory by Srednicki \cite{Srednicki:1993im} (see also \cite{Bombelli:1986rw,Casini:2009sr}). 

Indeed, in our case employing Srednicki's calculation is straightforward because the action in 
Eq. (\ref{action}) is quadratic when
$g_6 = 0$.  We can split the spatial $S^2$ 
into two regions, for example by taking a latitude circle and considering its two sides as separate regions, and then computing 
the entanglement in the full theory ground state on one side of the circle if we integrate out the modes residing on the other side.
The calculation is straightforward since the quadratic theory can be diagonalized. A subtlety is that
the theory must be regulated in the UV, which can be done by Fourier-tranforming on the azimuthal angle $\chi$, and 
discretizing the polar angle $\theta$ along a longitude. The UV cutoff then is the inverse distance $a$ 
between adjacent points on the longitude. This yields the result that 
$S \simeq \frac{\cal L}{a} \sim M_{\rm UV} {\cal L}$, where ${\cal L}$ is the latitude perimeter. By spherical 
symmetry, this quantity is maxed out when the latitude is taken to be a great circle on $S^2$ -- i.e. the equator. 

When gravity is turned on, and the UV cutoff $M_{\rm UV} \sim 1/a$ is taken to be the Planck scale, 
it has been forcefully argued \cite{Banks:2000fe,Banks:2001yp}
that this result is the basis dimension of the holographic dual of the bulk QFT coupled to 
gravity. In our case, we can add topological $2+1$ gravity 
\cite{Deser:1983tn,Deser:1983nh,Achucarro:1986uwr,Witten:1988hc} as a diagnostic probe, similarly to 't Hooft's 
use of Planck scale as a regulator \cite{tHooft:1984kcu} 
in the brick wall approach to black hole entropy\footnote{This ideology was also
used to reproduce the Bekenstein-Hawking bound in $2+1$ de Sitter \cite{Kim:1998zs}. A 
more refined method was discussed in \cite{Maldacena:1998ih}.}. This
brings about a dimensional scale -- the $2+1$ Planck scale, which serves the role of the ultimate UV cutoff. 
Its presence provides an upper limit to the energies and masses in the theory because the gravitational
effects in $2+1$ are global, instead of local. A mass in $2+1$, by virtue of peculiarities of $2+1$ gravity -- i.e., the absence 
of the Weyl tensor -- does not induce any local force, and leaves the surrounding space locally flat. However, its effects
persist arbitrarily far away, since a source induces a deficit angle. In order to allow a static
locally flat Minkowski space, for example, a mass at rest cannot exceed the Planck scale: more precisely, $M < M_3/4$, where
$M_3 = 1/G_3$ is the $2+1$ Planck scale \cite{Deser:1983tn,Deser:1983nh}. 

In our case this restricts the maximal energy allowed in our QFT on static Einstein. 
For ultrarelativistic modes, which are almost lightlike, we can implement the bound by considering the mode's 
gravitational field which is approximated by a $2+1$ variant of the Aichelburg-Sexl shockwave \cite{Aichelburg:1970ef}. 
We can obtain the solution in the flat space limit from the $AdS$ shockwaves 
given in \cite{Sfetsos:1994xa,Anber:2007ry}. This is a good approximation for the wave on $S^2$ at short distances. 
The solution is, with our convention that $G_3 = 1/M_3$, 
\be
ds^2 = 4du \, dw + dz^2 - 16\pi \frac{{\cal E}}{M_3} |z| \delta(u) du^2 \, .
\label{shock}
\ee
This solution can be attributed to a $\delta$-function source in the $2+1$ Einstein's equations, which are
$R^{\mu}{}_\nu = 8 \pi G_3 \, T^\mu{}_\nu$, and where $T^\mu{}_\nu \propto {\cal E}$. This means, the corrections
to the classical equations on this background are controlled by the expansion parameter $\zeta \simeq G_3 {\cal E}$, 
which therefore must be small for the classical regime to be valid; hence $\zeta < 1$, and so we find the bound
${\cal E} < M_3$ holds parametrically for relativistic modes too.  

Hence the highest energy excitation safely described by the effective theory cannot exceed in energy the critical value 
${\cal E}_{\ell~crit} \la M_3$. Since ${\cal E}_\ell =(\ell+1/2)H_0$, 
this means that the highest frequency below the cutoff are those with
\be
\ell_{crit} \la \frac{M_3}{H_0} \, .
\label{critl}
\ee
The total number of nonlocal modes described by zonal harmonics $P_\ell$ below the cutoff is 
\be
{\#_{\tt Hilbert}} \sim \sum^{\ell_{crit}}_{\ell=0} 1 \sim \ell_{crit} \simeq \frac{M_3}{H_0} \, .
\label{dimension}
\ee
This number scales as the proper 
perimeter of the largest circle on $S^2$ -- the equator -- which plays the role of the holographic screen in 
static Einstein. In this spacetime the equator is the hypersurface of maximal geodesic expansion, i.e. the apparent horizon 
\cite{Bousso:1999cb}. 
Any other hypersurface has a smaller perimeter. 
Thus this shows that the dimension of the nonlocal basis of de Sitter 
Hilbert space in the bulk of static Einstein is in fact set by the Bekenstein-Hawking 
bound ${\#_{\tt Hilbert}} \sim {\cal L}_{AH}/G_3$,
where ${\cal L}_{AH}$ is the perimeter of the apparent horizon. Thus, this counting parametrically 
produces a result that agrees with the entanglement entropy counting when the cutoff is the Planck scale, as argued in
 \cite{Banks:2000fe,Banks:2001yp}. 
 
The numerical value of the bound (\ref{dimension}) is identical to the horizon bound in $2+1$ 
de Sitter, even though the de Sitter 
horizon and the static Einstein equator hypersurfaces are in general different. 
Nevertheless, these hypersurfaces intersect at the horizon bifurcation point, i.e. at the de Sitter bounce. 
At the intersection, they coincide, being the apparent horizon simultaneously for both static Einstein and 
de Sitter. The value of the perimeter in Planck units constrains the Hilbert space dimension, and for a unitary theory 
the bound should stay the same -- as it does. While it is unclear 
how to directly relate the dual holographic theories beyond this point (it has been argued that de Sitter
time evolution could be interpreted as the dual RG flow \cite{Leblond:2002ns,Leblond:2002tf}), 
at least the bulk Hilbert space dimension remains unchanged by this, indicating unitarity at work. 

Bearing this in mind, Eq. (\ref{dimension}) shows that the dimension of 
the Hilbert space with the Planck scale as the UV cutoff is parametrically equal to the 
Bekenstein-Hawking entropy formula in $2+1$ static Einstein and de Sitter horizon:
\be
{\#_{\tt Hilbert}} \la S_{BH} = \frac{{\cal L}_{AH}}{4G_3} = \frac{\pi}{2} \frac{M_3}{H_0} \, .
\label{behawk}
\ee
This fits well with the ideas about the dimension of the Hilbert space of a QFT in
de Sitter being bounded by the 
Bekenstein-Hawking formula \cite{Banks:2000fe,Banks:2001yp}. While we do not have 
a detailed microscopic dual theory, we have found a basis of bulk modes which satisfies the bound.

\section{Summary}

In this paper, we have studied the conformally coupled scalar field theory interacting with external sources on $2+1$-dimensional
static Einstein universe $R \times S^2$. Since the theory is maximally symmetric on the background, it is exactly solvable with a discrete spectrum in any conformal gauge with the fixed finite radius of $S^2$. We constructed the Hilbert space of the theory using the standard 
canonical second quantization methods, having found the operator algebra and the corresponding Hilbert space basis states which are
Hamiltonian eigenfunctions. 

The energy spectrum of the Hamiltonian is very interesting, since it has a unique vacuum state, and excited states
whose energies are proportional to odd multiples of the energy 
gap given by the inverse radius of the sphere. Specifically, the local basis functions
are the spherical harmonics $Y_{\ell m}$ on $S^2$, and their time evolution is just a phase shift, like in flat space quantum mechanics and
quantum field theory. 
Due to this, the basis, and also any linear combination of it which represents an arbitrary state from the Hilbert space, 
display the phenomenon of full quantum revival: as time goes on,
the states evolve in such a way that any observable in any state repeats its value after a time interval $\Delta \eta = \pi \rho$. This means
that after we pick any time interval of width $\Delta \eta = \pi \rho$ on $R \times S^2$, the rest of the evolution is completely redundant
since the values of all observables regularly go back and forth between those at the endpoints. As a result, the theory is secretly 
a quantum field theory on de Sitter space, since the restriction to $I \times S^2$ is conformally isomorphic to $2+1$ de Sitter space.
This provides a way around the de Sitter horizons obstructing the standard formulation of QFT. 

When we introduce $2+1$ gravity, it provides the UV cutoff for the QFT, which limits the number of QFT states 
available to describe dynamics below the cutoff. The symmetries of the theory then show that 
among the basis modes, the dynamics -- i.e., the momentum transfer -- only proceeds through the combinations 
$P_\ell(\hat n \cdot \hat n')\propto \sum_m Y_{\ell m}(\hat n) Y_{\ell m}^*(\hat n')$,  when
probed by external sources which interact via scalar exchange. This recombines the contributions of the $2\ell +1$ 
degenerate states $Y_{\ell m}$ to only one relevant combination, $P_\ell$, given fixed locations of the sources. 
As a result, the total number of active dynamical modes 
is built from the basis of $P_\ell$'s, which for a given cutoff $M_3$ are ${\#_{\tt Hilbert}} \simeq \frac{M_3}{H_0}$. Parametrically, this is 
in excellent agreement with the Bekenstein-Hawking horizon bound in de Sitter, advocated 
to represent the dimension of de Sitter Hilbert space \cite{Banks:2000fe,Banks:2001yp}. 
It would be interesting to see if this can be extended to the case of nonlinear scalar 
field self-interactions, which we have not pursued here. 

While our example is rather special, relying on a large degree of symmetry, we find it interesting that it at least represents a consistent 
QFT on de Sitter, despite horizons. Perhaps some of the ideas discussed here might be used in the quest
for more examples. 

\vskip0.3cm

{\bf Acknowledgments}: 
NK is supported in part by the DOE Grant DE-SC0009999.

\end{document}